\documentclass[letterpaper,twocolumn,10pt]{article}
\usepackage{usenix2019}

\usepackage{epsfig,endnotes}
\usepackage{color,soul} % for highlighting.
\usepackage{hyperref} % for urls
\usepackage{hyperref}
\hypersetup{colorlinks=true,citecolor=blue,linkcolor=blue}
\usepackage{amsmath,amsopn,amssymb}
\usepackage{subfigure}
\usepackage{endnotes,microtype,xspace,graphicx,fancyvrb,multirow}
\usepackage{booktabs}
\usepackage{array,underscore,relsize}
\usepackage{tabularx}
\usepackage{comment}
\usepackage{soul} % strike out \st{Hello world}
\usepackage{listings}
\usepackage{fancyhdr}
\usepackage{xspace}
\usepackage{enumitem}
\usepackage[frozencache]{minted}
\usepackage[labelfont=bf,font=small,skip=5pt]{caption}
\usepackage[table]{xcolor}

\usepackage{fp}
\usepackage{siunitx}

\usepackage{balance}
\usepackage{flushend}
\usepackage[compact]{titlesec}
\usepackage{makecell}

\newcommand{\cc}[1]{\mbox{\smaller[0.5]\texttt{#1}}}

\newcommand{\metric}[1]{\multicolumn{2}{c|}{\bf #1 }}

\newcommand{\metricLast}[1]{\multicolumn{2}{c}{\bf #1 }}

\def\Snospace~{\S{}}
\newcommand{\sys}{\mbox{\textsc{MultiK}}\xspace}
\newcommand{\dkut}{\mbox{\textsc{D-Kut}}\xspace}
\newcommand{\skut}{\mbox{\textsc{S-Kut}}\xspace}
\newcommand{\kernelzero}{\mbox{\textsc{Kernel0}}\xspace}

\newcommand{\linuxversion}{4.4.1~}

\newcommand{\facechange}{\mbox{FACE-CHANGE}\xspace}
\newcommand{\kasr}{\mbox{KASR}\xspace}

\if 0

\fi

\newif\ifdraft\drafttrue
\newif\ifnotes\notestrue
\ifdraft\else\notesfalse\fi

\newcommand{\AU}[1]{\textcolor{purple}{AUSTIN: #1}}
\newcommand{\YJ}[1]{\textcolor{blue}{YJ: #1}}

\newcommand{\RB}[1]{\textcolor{red}{RBB: #1}}

\input{glyphtounicode}
\newcommand{\squishlist}{
	\vspace{10px}
	\begin{itemize}[noitemsep,nolistsep]
		\setlength{\itemsep}{-0pt}
		      }
		      \newcommand{\squishend}{
	\end{itemize}
	\vspace{10px}
}

\newcommand{\summary}{
	\begin{shaded}
		\vspace{-10px}
		\squishlist
		}

		\newcommand{\summaryend}{
		\squishend
		\vspace{-10px}
	\end{shaded}
}

\def\discretionaryslash{\discretionary{/}{}{/}}
{\catcode`\/\active
\gdef\URLprepare{\catcode`\/\active\let/\discretionaryslash
\def~{\char`\~}}}%
\def\URL{\bgroup\URLprepare\realURL}%
\def\realURL#1{\tt #1\egroup}%

\newcommand{\redcell}{\cellcolor{red!25}}
\newcommand{\greencell}{\cellcolor{green!25}}

\usepackage{tikz}
\newcommand*\CC[1]{%
	\begin{tikzpicture}[baseline=(C.base)]
		\node[draw,circle,inner sep=0.2pt](C) {#1};
	\end{tikzpicture}}

\usepackage{xstring}
\newcommand{\PP}[1]{
	\vspace{5px}
	\noindent{\bf \IfEndWith{#1}{.}{#1}{#1.}}
}

\newcommand{\PPC}[2]{
	\vspace{5px}
	\noindent{{\bf #1} (#2):}
}

\newcommand{\PC}[2]{
	\vspace{5px}
	\noindent{\bf\CC{#1} \IfEndWith{#2}{.}{#2}{#2.}}
}

\newcommand*\BC[1]{%
	\begin{tikzpicture}[baseline=(C.base)]
		\node[draw,circle,fill=black,inner sep=0.1pt](C) {\textcolor{white}{#1}};
	\end{tikzpicture}\hspace{2px}}

\newcommand{\KB}{\,\text{KB}\xspace}
\newcommand{\MB}{\,\text{MB}\xspace}

\newcommand{\etal}{\textit{et al}. \xspace}
\newcommand{\ie}{\textit{i}.\textit{e}., \xspace}
\newcommand{\eg}{\textit{e}.\textit{g}.,\xspace}
\newcommand{\etc}{\textit{etc}.\xspace}

\newcommand{\boxbeg}{
	\vspace{5px}
	\noindent\begin{tabular}{|l|}\hline
		\begin{minipage}{6.3in}
			\vspace{5px}
			\noindent
			}

			\newcommand{\boxend}{
			\vspace{5px}
		\end{minipage} \\ \hline
	\end{tabular}
	\vspace{1px}
}

\newcommand{\rbeg}{
	\begin{tabular}{|l|}\hline
		\begin{minipage}{6.3in}
			\vspace{5px}
			\noindent\textbf{Research Challenge.} \it
			}

			\newcommand{\rend}{
			\vspace{5px}
		\end{minipage} \\ \hline
	\end{tabular}
}

\newcommand{\ci}{{\it (i) }}
\newcommand{\cii}{{\it (ii) }}
\newcommand{\ciii}{{\it (iii) }}
\newcommand{\civ}{{\it (iv) }}
\newcommand{\ca}{{\it (a) }}
\newcommand{\cb}{{\it (b) }}
\newcommand{\ccc}{{\it (c) }}

\input{code/fmt}
\newcommand{\kconfig}{\emph{kconfig}\xspace}
\newcommand{\granularitycaption}{
    \textbf{B}, \textbf{S} refer to kernel profiles generated by \dkut with
    basic-block and symbol granularities respectively.
    \textbf{SC} refers to kernel profiles generated by \skut.
}
\newcommand{\granularitycaptionWoSc}{
    \textbf{B}, \textbf{S} refer to kernel profiles generated by \dkut with
    basic-block and symbol granularities respectively.
}

\begin{document}

%don't want date printed
\date{}
%make title bold and 14 pt font (Latex default is non-bold, 16 pt)
\title{\sys: A Framework for Orchestrating Multiple Specialized Kernels}
% \author{
	% \IEEEauthorblockN{Hsuan-Chi Kuo\IEEEauthorrefmark{1},
		% Akshith Gunasekaran\IEEEauthorrefmark{2},
		% Yeongjin Jang\IEEEauthorrefmark{2},
		% Sibin Mohan\IEEEauthorrefmark{1},
		% Rakesh B. Bobba\IEEEauthorrefmark{2},
		% David Lie\IEEEauthorrefmark{3}, and
		% Jesse Walker\IEEEauthorrefmark{2}
	% }

	% \IEEEauthorblockA{\IEEEauthorrefmark{1}Dept. of Computer Science, University of Illinois at Urbana-Champaign, USA \\
		% \{hckuo2, sibin\}@illinois.edu}

	% \IEEEauthorblockA{\IEEEauthorrefmark{2}Dept. of Electrical Engineering and Computer Science, Oregon State University, USA \\
		% \{gunaseka, yeongjin.jang, rakesh.bobba, walkjess\}@oregonstate.edu}

	% \IEEEauthorblockA{\IEEEauthorrefmark{3}Dept. of Electrical and Computer Engineering, University of Toronto, Canada \\
		% lie@eecg.toronto.edu}

% }
\author{
    {\rm Hsuan-Chi Kuo}\\
    University of Illinois at Urbana-Champaign
    \and
    {\rm Akshith Gunasekaran}\\
    Oregon State University
    \and
    {\rm Yeongjin Jang}\\
    Oregon State University
    \and
    {\rm Sibin Mohan}\\
    University of Illinois at Urbana-Champaign
    \and
    {\rm Rakesh B. Bobba}\\
    Oregon State University
    \and
    {\rm David Lie}\\
    University of Toronto
    \and
    {\rm Jesse Walker}\\
    Oregon State University
    % copy the following lines to add more authors
    % \and
    % {\rm Name}\\
    %Name Institution
} % end author

\maketitle

% Use the following at camera-ready time to suppress page numbers.
% Comment it out when you first submit the paper for review.
% \thispagestyle{empty}

\begin{abstract}

    %\hl{current version -- Sibin; Feb 15th 2:45 AM CT}\\
    We present, \sys, a Linux-based framework~\footnote{This concept is
    generalizable to any operating system that uses paging.} that reduces
    the attack surface for operating system kernels by reducing code bloat.
    \sys ``orchestrates'' multiple kernels that are
    \textit{specialized} for individual applications in a transparent manner.
    This framework is flexible to accommodate
    different kernel code reduction techniques and, most importantly,
    run the specialized kernels with
    \textit{near-zero additional runtime overheads}.
    \sys avoids the overheads of virtualization and runs natively on the system.
    For instance,
    an Apache instance is shown to run on a kernel that has
    \ca $93.68 \%$ of its code reduced,
    \cb $19$ of $23$ known kernel vulnerabilities eliminated and
    \emph{(c)} with negligible performance overheads ($0.19 \%$).
    \sys is a framework that can integrate with existing code reduction and
    OS security techniques.
    We demonstrate this by using \dkut and \skut --
    two methods to profile and eliminate unwanted kernel code.
    The whole process is transparent to the user applications because
    \sys \textit{does not} require a recompilation of the application.

\end{abstract}

\section{Introduction}
\label{s:intro}

General-purpose operating systems (OSes) have, over time, bloated in size.
While this is necessitated by the need to support a diverse set of applications
and usage scenarios, a significant amount of the kernel code is typically not
required for \textit{any given application}. For example, Linux supports more than $300$ system calls and contains code for supporting different filesystems, network protocols, hardware drivers, \etc all of which may not be needed for every application or deployment. While a minimal
off-the-shelf install of Ubuntu 16.04 (running kernel 4.4.1) produces a kernel
binary with an $8 MB$ text section, many of the applications that we profiled
(refer to \autoref{s:eval} for more details) only use about $800 KB$ of it.

In addition to performance issues, unused kernel code (when mapped into an
application's process memory) represents an attack surface -- especially if vulnerabilities exist in the unused parts of the kernel code. Such vulnerabilities, while less common than those in
applications, are still found with regular frequency.\footnote{Over $2500$
vulnerabilities have been found in the Linux Kernel since 2010 (\url{
https://nvd.nist.gov/}).} Since OS kernels are often considered to be a part of
the trusted computing base (TCB) for many systems, this attack surface poses a
significant risk. Today, there exist many known exploits that take advantage of
kernel vulnerabilities
(\eg~CVE-2017-16995\footnote{\url{https://access.redhat.com/security/cve/cve-2017-16995}}).

%Recognizing the bloat in operating system kernels and the increased attack surface it represents
Researchers have explored different techniques to reduce kernel code (\eg ~\cite{unikernels,Tartler-HotDep12,kurmus,Face-Change-DSN14,kasr}). For example, \ca building application specific \emph{unikernels}~\cite{unikernels},
\cb tailoring kernels through build configuration editing~\cite{Tartler-HotDep12,kurmus},
\cc providing specialized kernel views for each application~\cite{Face-Change-DSN14,kasr} among others. These approaches, either need application level changes~\cite{unikernels}, or need expert knowledge about (and manual intervention in) the selection of configurations -- they also sacrifice the amount of kernel reduction achieved to support multiple applications~\cite{Tartler-HotDep12,kurmus}, or incur significant performance overheads~\cite{Face-Change-DSN14} or can only specialize the kernels at a coarse page level granularity ~\cite{kasr}. Note: ``granularity'', in this context, refers to sizes of code chunks that are considered for elimination; some techniques eliminate kernel code at the page level~\cite{kasr} while others may choose to do it at a basic block level~\cite{Face-Change-DSN14}. %\footnote{In this paper we show that it can also be carried out at a finer granularity of basic blocks (\autoref{s:kern-speci}).}.
%\RB{I still think the footnote and the following sentence are not helping us}
We show that our framework can evaluate systems with different levels of code reduction granularity (with the obvious result that with a finer granularity of code reductions, a greater amount of kernel code can be eliminated (\autoref{s:sec-eval})).
%, thus demonstrating the power of our framework -- the ability to carry out system-level evaluations of different ideas.
%$$\RB{While code reduction at finer levels of granularity can lead to better attack surface reduction, in previous work such fine grain reduction came with higher overhead costs~\cite{Face-Change-DSN14}}.

% On the other hand, removing Kernel code could result in an unstable system due to the fact that
% some functionality is lost during the process. Specializing the system for one application
% could also render it unusable for another. In certain cases it is just not worth compromising
% on stability for reducing the trusted code base; thus the entire
% kernel should be made available to the application.
% There is also a need to support more than one application on the same system
% either due to dependencies or to save cost.
% Hence our objective is:
% \ci limit the application's "view" of the kernel to code regions that are
% required by the application
% \cii is isolated from other applications "view" while also,
% \ciii ensuring that each application has enough code to run as expected without instabilities.

\begin{figure}[hb]
	\includegraphics[width=\columnwidth]{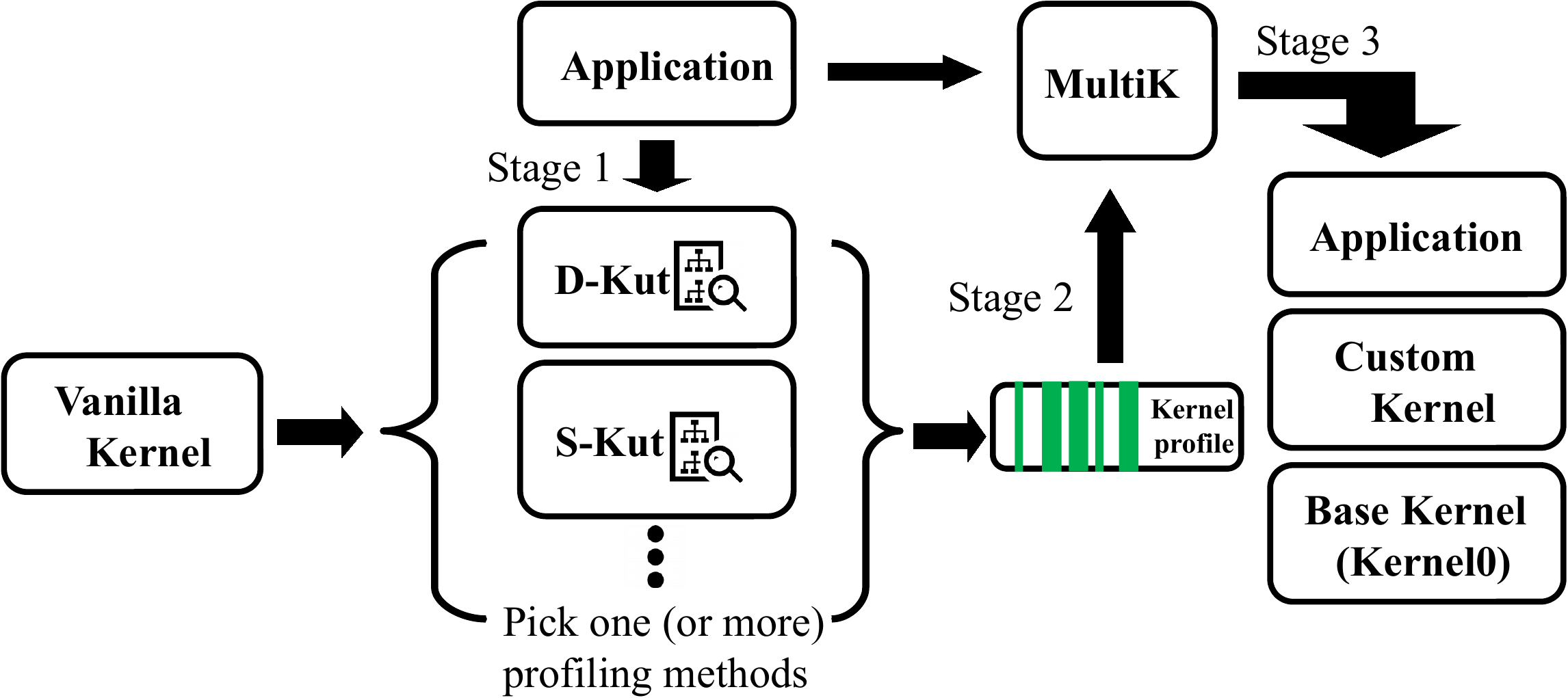}
	\caption{High level architecture of \sys.
        \textbf{Stage 1}: an application is profiled to identify necessary kernel code using techniques chosen by developers (this paper
        presents two such possible techniques: \dkut and \skut).
        \textbf{Stage 2}: \sys reads the information for code reduction
        (based on the kernel profile) generated in Stage 1.
        \textbf{Stage 3}: \sys creates the customized kernel (based on kernel profile from in Stage 2) when the application is launched.
	}
	\label{fig:highlevel-arch}
	\vspace{-\baselineskip}
\end{figure}
In this paper we introduce
\sys (\autoref{s:overview}),
\textit{a flexible, automated, framework for \ca orchestrating multiple
kernels (full or specialized) at \cb user chosen code reduction granularity
and \cc near-zero runtime overheads} -- without the need for virtualization.
    %\footnote{Granularity
    %refers to the aggressiveness in terms of code reduction. For example, page level, symbol level, block level \etc},
%on a single system -- without the need for virtualization and near-zero runtime overhead}.
%\hl{Granularity of specialization refers to the size of code increments that are considered for elimination}.
%A more fine-grained specialization (\eg byte size vs. page size) results in better attack surface reduction.
In addition, \sys 
%can support multiple granularity of specialization, and more importantly, 
does not introduce additional overhead if a finer-grain granularity is chosen. To demonstrate this, we evaluate \sys on three levels of granularity using two specialization tools (\autoref{s:kern-speci}),
\ci \dkut, a run-time profiling-based tool that can tailor the Linux kernel at different granularity (we perform basic block- and function/symbol-granularity tailoring in this paper)
and \cii \skut, a syscall-based tool that tailors the kernel based on the system calls used by the application.
\sys is able to simultaneously run multiple applications
that either \ci have their kernels specialized with any one of these tools or \cii use the full,
unmodified, kernel. Note that, though the two approaches (\dkut and \skut)
%\RB{which two apoproaches? S-KUT and D-KUT? }
are complementary -- they can either be used
independently or in conjunction with other tools/frameworks\footnote{In fact,
\sys is not beholden to these tools. System designers may use their favorite
profiling methods/tools. The kernel profiles that are obtained can be easily integrated with the \sys framework.}.

\autoref{fig:highlevel-arch} shows the high-level architecture of our \sys framework. A vanilla kernel is profiled (Stage 1) using the application that
is supposed to run on it. Note that in contrast to existing approaches,
the system designer can choose the level of granularity and code reduction
techniques (Stage 2). At runtime (Stage 3), the \sys framework launches the
application on its specialized kernel (one application per specialized kernel\footnote{Though this can be generalized as explained in \autoref{s:discus}.}).
The entire framework executes on a ``base'' (vanilla) kernel. The details of
each of these stages/components are explained in the rest of the paper.

\sys imposes very small, almost negligible, runtime overheads ($< 1\%$)
\emph{regardless of the granularity chosen} (\autoref{s:eval}).
User programs run unmodified and natively. In fact, the entire framework is
transparent to application developers. Our framework can also easily integrate with container-based systems such as Docker~\footnote{https://www.docker.com} (see our experiments in \autoref{tab:docker-perfeval}) -- which are popular for deploying applications.

%In contrast to previous approaches, when profiling an application (Stage 1 in \autoref{fig:highlevel-arch}),
%\sys allows the application deployer to choose the level of granularity and specialization
%technique (Stage 2).
%During runtime (Stage 3), \sys imposes a very small application startup overhead and
%virtually zero ($< 1\%$) runtime overheads \emph{regardless of the specialization techniques} (Section \autoref{s:eval}).
%Also, user programs can run unmodified, natively. In addition, the entire system is
%transparent to application developers. This framework can also easily integrate with container-based systems
%such as Docker~\footnote{https://www.docker.com} (see our experiments in \autoref{tab:docker-perfeval}) which are a popular way to deploy applications.

In summary, this paper makes the following contributions:
\begin{enumerate}
  \item Design and implement \sys (\autoref{s:multik}), a flexible \emph{framework} for dispatching
      multiple kernels (specialized, unmodified or a combination of both) on the same machine without the need for virtualization.
  \item Build \dkut (\autoref{s:dkut}) and \skut (\autoref{sec:syscallanalysis}), tools for specializing a kernel
      at fine (instruction-level) and coarse (system-call level) granularity, respectively.
      %\item a \emph{complementary kernel specialization technique} that is granular \RB{should we highlight it here? Face-change does the same thing right?}
  \item Presents an evaluation for attack surface reduction and performance overheads;
      the proposed framework shows virtually no runtime overhead
      while significantly reducing the attack surface
      both in terms of the amount of eliminated code and CVEs.
\end{enumerate}
We also discuss limitations of \sys and also how other techniques can be adapted to integrate with it (including a comparative evaluation) in \autoref{s:eval} and \autoref{s:discus}.

\section{Background and Design Goals}
\label{s:overview}
Kernel specialization for attack surface reduction has been studied in multiple previous works.
Those prior works all aimed to \ci identify unused parts of a bloated kernel
and \cii remove the identified parts either statically or disabled them at runtime.
Debloating kernels in this way removes
software vulnerabilities present in unused kernel code, reducing its attacks surface and thereby contributing to a system's security.
In the following,
we address limitations of existing
kernel attack surface reduction approaches and then
set our threat model and design goals.
At the end,
we illustrate an overview of our approach
in \sys to achieve our design goals.

\subsection{Limitations of existing approaches}

\PP{Complexity in handling Kconfig}
One of the prominent approaches to de-bloating Linux kernel is
to use the kernel build configuration system, \emph{kconfig}.
However, working with those configurations is a complex job
not only because there are
over 300 feature groups and more than 20,000 individual
configuration options to choose from,
but also because most of these options have many dependencies that
further contribute to the complexity of the system.
While approaches to automate \emph{kconfig} to tailor
the Linux kernel have been proposed~\cite{Tartler-HotDep12,kurmus},
they often require (manual) maintenance of
whitelist and blacklist configuration options
--- these lists quickly become irrelevant as applications and kernel evolve.

\PP{Specializing a kernel for running many applications}
When running multiple application on a system, those approaches~\cite{Tartler-HotDep12,kurmus} tailor the kernel for
the \textit{combined} set of applications.
%
%
%\RB{Note to myslef: Need to modify the rest of this paragraph.
%Isolation argument is weak for us and we shouldn't highlight it I think.}
%
%This results in \ca not much code reduction to begin with and
%\cb the lack of isolation between applications.
%
This negatively affects the attack surface reduction because
the kernel will need to contain code for serving requests from
all such applications, and not just for serving a single application.
For instance,
orthogonal applications (\eg Apache and ImageMagick) only share about
$20\%$ of the kernel code for their usage.
Even similar applications (\eg Apache and \cc{vsftpd}) share about
$83\%$ of the kernel and often leave as much as $20\%$ not shared.
Hence, there is a high likelihood that unused code stays in the final kernel.

\PP{Use of virtualization and specialization granularity}
To address some of these concerns,
the Face-Change system proposed the customization of kernels for
each application~\cite{Face-Change-DSN14}.
However, their system is implemented with a hypervisor component,
and this makes
applications and their kernels run in a virtual machine
(with associated performance penalties).
Further,
due to the use of a VMI-based approach for
determining the appropriate kernel "view",
Face-Change incurs additional runtime overheads
(about $40\%$ for Apache and around $5 - 7\%$ for Unixbench).

KASR system~\cite{kasr} eliminates the performance overhead
(keeps it within $1\%$)
but still requires applications to run in virtual machines.
Moreover,
their kernel specialization is limited to a coarse
page-level granularity
that can still allow unnecessary and potentially vulnerable kernel code to remain in the system. With \sys we aim to overcome some of these limitations as we discuss in~\autoref{subsec:goals}.

%\RB{DL text begin}
\PP{Application kernel profile completeness.}
A precondition of using any of these kernel reduction systems is
an accurate and complete profile of
the kernel facilities that applications depend on.
If this profile is incomplete,
then a benign, uncompromised application may
try to execute code in the kernel that is not part of the profile.
Unfortunately,
executing code that isn't in the customized kernel
due to an incomplete profile is indistinguishable from
a compromised application trying to execute code that
the original application cannot invoke.
The need for a complete profile is
a limitation of all kernel reduction systems~\cite{Tartler-HotDep12,kurmus,Face-Change-DSN14,kasr}, including \sys.
%DL text end

%In this section,
%we describe the design goals for \sys and
%provide a high-level overview of our approach.
%
%Additionally, we point out limitations of existing systems.

\subsection{Threat Model}
    In \sys, we assume the following for attackers:
    \squishlist
      \item \PP{Local, user-level execution is given
      without physical access to the machine}
      We assume our attackers are limited to launching
      local/remote attacks on kernel from user privilege (\ie ring 3)
      without having any physical access to the machine.

      \item \PP{Firmware, BIOS and processor are trusted}
      Attacks on the kernel originating from components
      at levels lower than the kernel are out of scope for this paper.

      \item \PP{Hardware devices attached to the machine are trusted}
      Similarly, DMA attacks and other attacks from hardware devices are
      out of scope.
    \squishend

This threat model covers general kernel exploit scenarios, \ie launching an attack from user-level to interfere with kernel-level execution.
For instance,
the followings are examples of valid attacks on \sys: %\hl{TODO}

    \squishlist
    \item Privilege escalation attacks from user to kernel.
    \item Control-flow hijacking attacks (arbitrary code execution) in kernel.
    \item Information leaks (arbitrary read) from kernel to user.
    \item Unauthorized kernel data overwriting (arbitrary write)
    originating from user.
    \squishend

\subsection{Design Goals}\label{subsec:goals}
The overarching goal of \sys is
to reduce the kernel attack surface of a system
by generating a minimal kernel for running
a set of user applications.
To achieve this goal,
we aim to build  a system that can do this in
an efficient (\ie~no runtime overhead) and
transparent (\ie~no application changes) manner.
We elaborate on the design goals of \sys next.

\PP{Flexible and fine-grained attack surface reduction}
The first design goal of \sys is
to only permit the kernel code identified as necessary in an application's profile to run when the application is running.
Some prior work~\cite{kurmus} customizes kernels at \emph{feature} granularity,
by minimizing build-time configuration options
required for supporting a specific application
(\eg if access to USB mass storage is not required,
\cc{CONFIG_USB_STORAGE} can be disabled to exclude code
that deals with interfacing with USB devices).
Sometimes such features are big,
and because the feature is its minimal granularity,
it often contains more code than required
even if only parts of features are used by applications.
Similarly, KASR~\cite{kasr} customizes kernels at a \emph{page} granularity
by dynamically removing the executable permission for
specific kernel code pages if
the code in those page are not being used by the application.
However, this approach overestimates required kernel code by including
a whole page (4~\KB) even if application requires only parts of it.
%DL edits
In this regard, we aim to design \sys to
support granularity down to the basic block level.  We note that in theory, \sys can even support byte-level granularity, but in practice since it would not make sense on current CPU architectures to permit a subset of instructions in a basic block to execute without permitting all the instructions in the basic block to execute, we only evaluate \sys down to the basic block granularity.
%DL edits end
Further, prior works often could only customize at a specific granularity (\eg page level for KASR~\cite{kasr}, feature level for Tailor~\cite{kurmus}). Our goal is to design \sys
to be a framework that can orchestrate kernels specialized at different specialization granularity.

%\RB{Are we doing instruction level?}\AU{Block level only but like what yeongjin said we can mention that}(even the instruction-level granularity shown in \autoref{sec:dkut}).

\PP{Fine-grained security domain for customization}
Another design goal for \sys is
to customize kernel at a fine-grained security domain level.
A security domain can be a single process or
an instance of a container.

Previous works customized the kernel for a whole system
(\ie all applications or
an application stack running on the machine)~\cite{kurmus} or
to run a whole virtual machine~\cite{kasr}.
Such customized kernels will include
the union of kernel code required
by every application running on the machine.
Customizing kernels for all applications together does not minimize the attack surface as
each application will have
more kernel code mapped than is required~\cite{Face-Change-DSN14}.
As a result, \sys aims to support specialized kernels for each
process so that every specialized kernel has code only necessary for
the process.

\PP{Efficiency}
\sys should minimize the performance overhead.
Specifically, we aim to have
near-zero (\textless~1\%) as shown in \autoref{subsec:perfeval} run-time performance overhead
and not interfere with the application execution.

\PP{Transparency}
\sys should not require application source code,
application code instrumentation or application changes.
Further,
customized kernels must be compatible with the target application
and should able to support regular use-cases or workload of the application.
To this end, we design \sys
to interact only with the kernel space
to maximize compatibility and
to support user-level applications transparently.

\PP{Flexibility in sharing system resources}
%\YJ{Required, and need a fix. }
Some applications work and interact with each other closely (\eg Apache + git for GitLab and Apache + ImageMagick for MediaWiki). For such applications \sys should allow applications to interact using system resources (\eg IPC, locks, \etc) to maximize the flexibility in application interactions.
Employing virtual machine based solutions
and address space isolation techniques (SFI~\cite{DBLP:conf/sosp/WahbeLAG93}, XFI~\cite{DBLP:conf/osdi/ErlingssonAVBN06}, \etc),
will reduce the flexibility and ease of such interactions.
Ideally, \sys should let normal interactions among applications
as if they are all running on a machine
with a single kernel.
Running multiple applications in \emph{docker} containers on the same machine
exemplifies this goal.
In other words, we would like our design to provide
customized kernels for each application (or docker container)
to strike a balance between the isolation of attack surfaces
and the flexibility (\ie to allow normal application interactions).

% \PP{Attack detection and graceful degradation}
%Kernel customization involves tailoring the kernel to have the minimum set of instructions required to support an application and removing the unused instructions.
%
% \sys should fail gracefully in case an application
% tries to access removed kernel code (either because of
% poor profiling during customization, due to changes in application workload or because of an unexpected execution caused by an attack).
% %
% In such cases, instead of crashing the kernel,
% \sys should report an anomaly and
% kill the application safely so as not to disturb
% other executions in the system.

\subsection{Our Approach}
%\YJ{Need to mention that \sys is orthogonal to kernel customization technique.
%\ie, any \emph{statically} customized kernel can be deployed by \sys\dots}
%

\begin{figure}[ht]
\vspace{-\baselineskip}
    \includegraphics[width=\columnwidth,keepaspectratio]{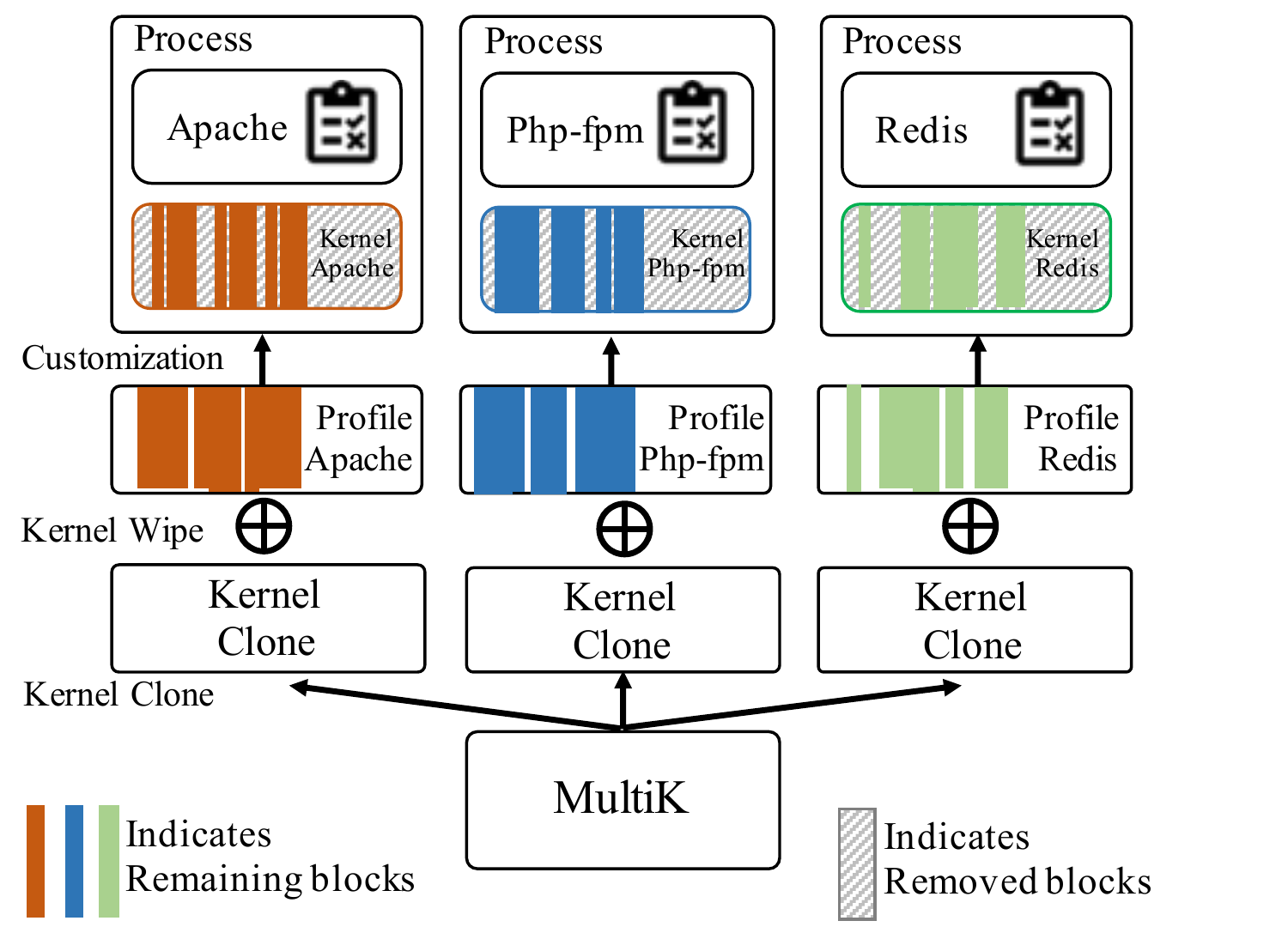}
    \caption{
        \sys Architecture.
        This figure depicts
        the runtime kernel specialization of \sys.
        When an application is launched,
        \sys first fetches the application's kernel profile
        from the kernel profile database.
        Next, \sys clones the \kernelzero as \textsc{Kernel0'}
        and then masks the parts that are not contained in
        the application's kernel profile.
        As a result, \sys creates a customized kernel for the running application.
        \sys applies this process for all applications running in the system,
        and each application runs with its own customized kernel.
    }
    \label{fig:multik}
\vspace{-\baselineskip}
\end{figure}

\PP{MULTIK: Multi-Kernel Orchestration} We present \sys (\autoref{fig:multik}), a kernel runtime that deploys
tailored kernel code for each application.
\sys customizes the kernel code
before launching an application by
consulting the \emph{kernel profile} of the target application.
Specifically,
\sys makes \ca a copy of the entire kernel text to
a new physical memory region,
\cb removes the unused parts from this kernel copy by
using the application's kernel profile
and \ccc deploys the application with this tailored kernel.
To transparently and efficiently deploy
tailored kernel code for each application,
page table entries for the kernel text are updated.
That is, \sys alters virtual-to-physical page mappings to
point the application to the new customized kernel.
By doing this,
we can switch automatically the kernel between applications (with
near-zero performance overhead)
because the page table will be switched
as part of a context switch.

\sys can work with application \emph{kernel profiles} produced at different granularities and using different profiling techniques. We use two specialization techniques (presented below) to demonstrate and evaluate \sys.

\PP{D-KUT: Dynamic tracing}
\dkut (\autoref{fig:kut}) is a tool to identify the kernel code necessary for an application.
We leverage dynamic tracing of the kernel execution while
the system is running the target application with its use cases.
This approach has also been used in previous work (\eg ~\cite{Tartler-HotDep12, kurmus,Face-Change-DSN14,kasr}).
The purpose of this step is
to determine which parts of kernel code to remove before application deployment.
In particular, we will remove any code that is not included
in the dynamic execution traces.
Our tracer collects and stores executed parts of the kernel at
basic-block granularity (the smallest granularity)
to maximize attack surface reduction.
For tracing to be effective and
for the customized kernel to not impact application execution,
tracing requires a good set of use cases that
are representative of an application's kernel usage. Previous work (\eg ~\cite{kurmus}) has found that a modest amount of test cases and tracing runs are sufficient to get good coverage of an application's kernel profile.
Here we assume that such test cases are supplied by
application developers or deployers
(\eg from unit tests, benchmarks, workloads \etc).
Generating the workloads or test cases is not in scope for this work
because \sys does not limit how the kernel is profiled.
Since different applications exercise different parts of the kernel,
we need to carry out the tracing for each application and
store them as an application's \emph{kernel profile}.
This collected kernel profile is used by \sys
when it generates a tailored kernel for an application at runtime.
\textbf{Note} that \sys works independently of
the kernel customization technique and
can use a kernel profile generated by
any existing or future customization techniques.

\PP{S-KUT: Syscall Analysis}
\skut is an approach to expand the kernel profile and enhance the reliability.
To be more specific, \skut tries to include rarely-triggered exception/ error handling
code while being able to remove a large portion of the kernel.
We use the compiler features to analyze the kernel source code.
By doing this, we can have precise information of what code should be included.
For this approach to be effective, the deployer has to have a list of system calls
that the application uses by either static (\ie symbolic execution)
or dynamic methods (\ie \cc{strace}). \sys does not address and limit
how the list of system calls is generated.

\PP{Benefits}
An immediate benefit of
removing large portions of unused kernel code is
the resulting attack surface reduction.
Vulnerabilities (both known and unknown) present in the removed code will never exist in the resulting runtime kernels.
%
%Thereby,
%systems running such kernels are free from the attacks
%originated by the removed vulnerabilities.
%
Our evaluation results show that
\sys successfully removes large portions of kernel code
(\eg unused kernel functions,
system calls and loadable kernel modules)
from the application's memory space.
In many instances we achieve more than $90\%$ reduction in kernel code,
\ie\cc{.text} (see~\autoref{tab:reduction} in \autoref{s:sec-eval}) and
consequently eliminate many vulnerabilities.
For instance,
\sys eliminates 19 out of total 23 vulnerabilities (listed as CVEs) in
Linux Kernel 4.4.1 when tailored for the Apache web server
(see \autoref{s:sec-eval}).
%
%Moreover, \sys also eliminates the potential  zero-day vulnerabilities stem from the removed parts of the kernel.
%
%Additionally,
%our performance evaluation on specialized kernels shows that
%the runtime performance overhead of \sys is
%less than 1\% (see \autoref{subsec:perfeval}),
%and can run applications such as Apache, Redis, gzip, STREAM, and perf
%transparently without any modification.
%
Although \sys does not aim to
detect and eliminate specific vulnerabilities,
it can reduce the system's overall risk of compromise
and could serve as one layer of a
\emph{defense-in-depth}~\cite{defense-in-depth} strategy.

\section{\sys: Orchestrating Specialized Kernels}
\label{s:multik}

\sys is a kernel mechanism to
orchestrate customized kernels efficiently and transparently
for each application at runtime.
The goal of \sys is to provide each application access to only kernel code that is customized to that application.
%\YJ{for given kernels customized per each application will run in the system---needs a different wording.. my bad..},
%
By doing so,
\sys will reduce the kernel attack surface by
removing a large part of unused kernel code
from the virtual memory of the application processes.

\begin{figure}[t]
	\includegraphics[width=\columnwidth]{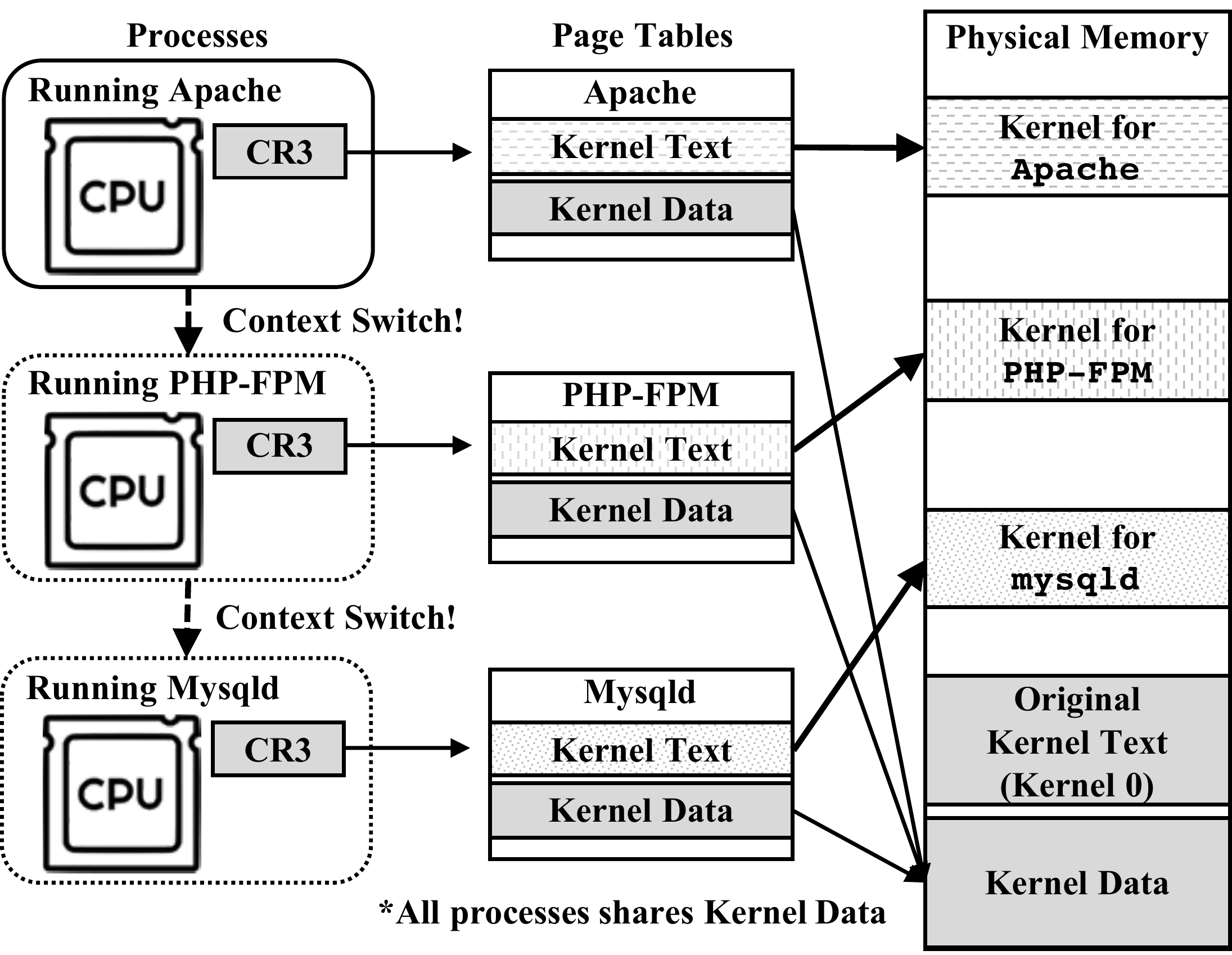}
	\caption{Kernel switching in \sys.
		To support an efficient and transparent kernel switch,
		we alter each application's page table (pointed by CR3)
		to map the virtual addresses for kernel text
		to its customized kernel code.
		In this way,
		switching among customized kernels is
		handled automatically by
		regular context switch in the kernel.
		That is,
		the page table will be switched automatically by the kernel
		during the context switch.
		Note that we only customize kernel code,
		so kernel data is shared among all processes.
	}
	\label{fig:switch}
\end{figure}

\PP{Overview}
\autoref{fig:switch} illustrates how \sys works with
multiple applications running on the system.
During the launch of each application,
\sys maps the customized kernel for the application into
a new physical memory region.
\sys then updates the page table entries (of the kernel code)
for that application process
to redirect all further kernel execution to
the customized kernel.
This update to the page table will completely remove
the original (full) kernel from
an application's virtual memory view
and switch the view to the customized one.
Additionally,
this update will guarantee that
the CPU that runs the application
can only work with the customized kernel code.
This is because the process context switching in the operating system
will switch the page table automatically.
Hence virtual addresses in the CPU will only refer to the customized view of the kernel code.

This approach is promising
because once the page table is updated during application launch (one-time cost),
no runtime intervention is required to switch to the customized kernel.
This will significantly enhance the runtime performance.
In contrast to this approach,
prior work relies on changing extended page table (EPT) permissions at runtime (in \kasr ~\cite{kasr}).
This has a limitation of page-level permission deprivation,
or relies on virtual machine introspection (VMI) to customize page table entries
at each context switch (in \facechange ~\cite{Face-Change-DSN14}).
All of which incur nontrivial runtime performance overheads.

\subsection{Challenges}

Although our design is conceptually simple,
\sys must overcome a few challenges:

\PPC{Sharing system resources}{\autoref{s:design-deploy}}
Running multiple applications in the same system could
benefit from resource sharing among the applications.
For instance,
processes can communicate efficiently via
inter-process communication (IPC) mechanisms available in the system
such as locks, signals, pipes, message queues, shared memory,
sockets, \etc
and processes can also share hardware devices attached to the system.
Such sharing is transparent by design as applications
share the same kernel (\ie code) and
share the same kernel memory space (\ie kernel data).
However, running a customized (a different) kernel per each application
could interfere with such transparent sharing.
For instance,
running a customized kernel in a virtual machine requires
virtualizing hardware and shared resources among applications
that introduces compatibility and performance issues.
Additionally,
having a different memory layout for each application's kernel
would make the data sharing even harder.
For instance, a data structure prepared in one application cannot be used
in a different application and might even require transformation.

\PPC{Handling hardware interrupts}{\autoref{s:design-interrupt}}
A hardware interrupt can occur at any time regardless of its customized kernel view.
For instance,
while running a customized kernel for a non-networking application
(such as \cc{gzip}), the CPU could encounter
a hardware interrupt from the network interface card (NIC).
In such a case, the execution will be redirected to
the interrupt handler. If the customized kernel is not equipped to handle the interrupt,
then the system could either become unstable or important
events could be missed.
An easy workaround would be to include
all hardware interrupt handlers in the customized kernel,
as \facechange~\cite{Face-Change-DSN14} does.
However, such an approach will increase the kernel attack surface
by adding unnecessary kernel code for an application
that doesn't need access to particular hardware.

% \RB{REMOVE THIS paragraph below?} \AU{+1 remove}
% \PPC{Executing missing code}{\autoref{s:design-tolerance}}
% %
% If aggressively customized, the kernel may not have
% all the required code for the target application's execution.
% %
% This could happen for example if the profiling methods do not provide good coverage
% of kernel code usage by the application as may be the case for complex applications.
% %
% Depending on how the customized kernel handles faults
% caused by missing kernel code, the system could halt
% or enter an unexpected state if not carefully handled.
% Both of these are not desirable when deploying systems that need to be dependable.

\subsection{Deploying Customized Kernels}
\label{s:design-deploy}

\begin{figure}[t]
\vspace{-\baselineskip}
	\includegraphics[width=\columnwidth]{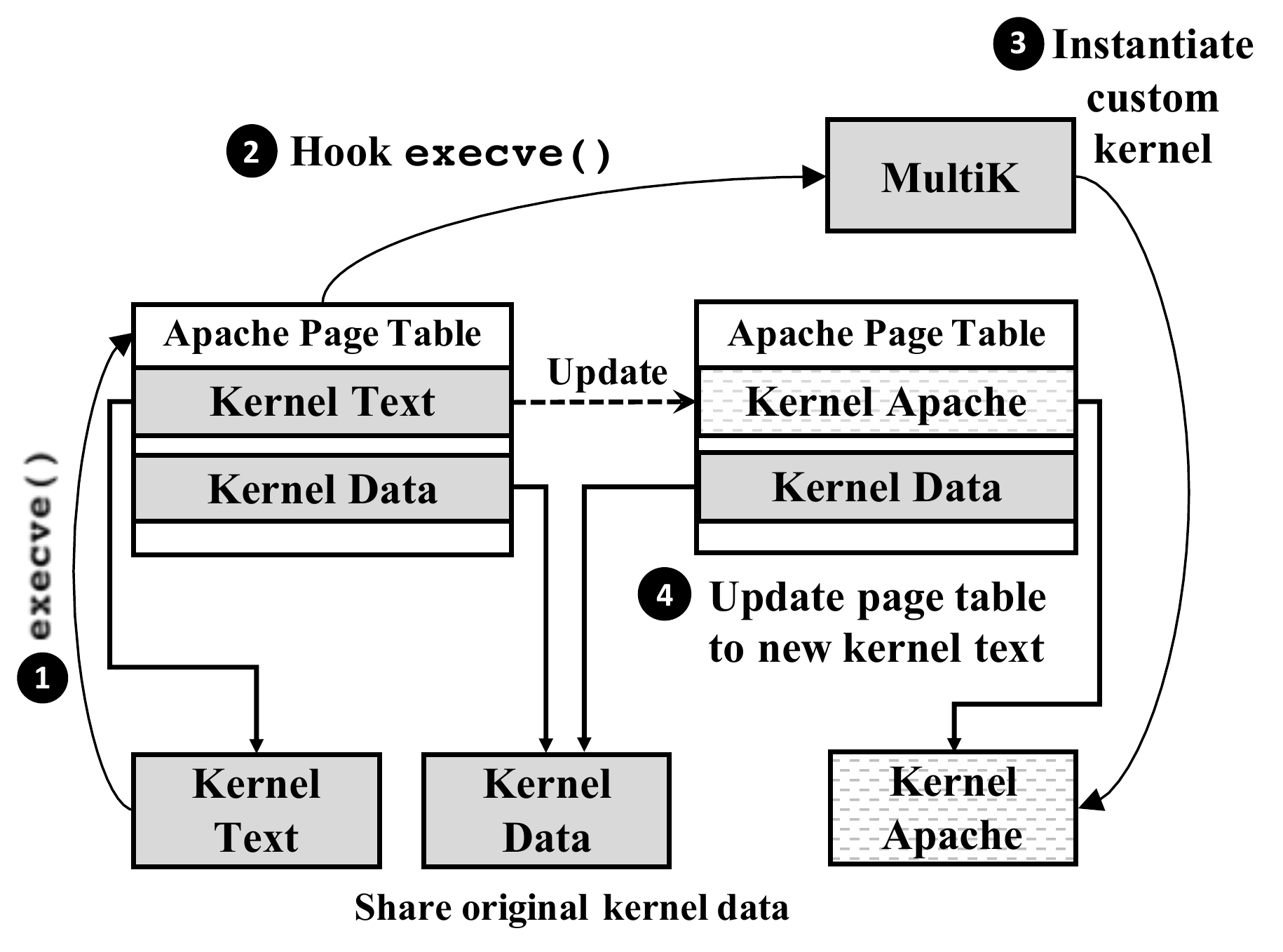}
	\caption{\sys's hook in \cc{execve}.
		On an application's execution,
		\sys will map the customized kernel for the application
		and then update the application's page table entries for
		kernel text to point this custom kernel.
	}
	\label{fig:execve}
\vspace{-\baselineskip}
\end{figure}

%\YJ{It's so simple so it's very difficult to write it in a good way...}

When deploying customized kernels in \sys,
applications should only have a view of the kernel customized for them.
At the same time,
we would like to ensure that
customized kernels running in one system could share
system resources, \eg share the memory space, available IPC mechanisms \etc
To achieve these goals in a manner transparent to the application,
we update an application's page table entirely
(corresponding to kernel text)
to point to the customized kernel
and we do this when launching the application
(\ie at \cc{execve()}).

Deploying the customized kernel by updating
kernel text page table entries gives us
the following benefits.
First,
sharing system resource among multiple kernels becomes straight forward.
Each application shares
the kernel data memory space
because we only alter the page table entries for kernel text.
Sharing system resources via
shared memory, direct memory access, or
sharing kernel data structures is transparent.
Second,
loading and switching between kernels is simple and efficient.
Loading a custom kernel will only incur the costs for mapping of new code section and updating page table entries,
which is simple to do as part of the \cc{execve} system call.
Additionally, switching between the kernels is simple
because a typical context switch involves
changing the current page table pointer (\cc{CR3} register in x86)
to that of the newly scheduled application,
and this will automatically switch the kernel text too.
In short,
the goal of providing applications access to only customized kernel code will be
met in a transparent manner.
Therefore, \sys always isolates the application's kernel view
to the one customized for it.

\autoref{fig:execve} (from \BC{1} to \BC{4})
shows how \sys switches the page table entries during application launch.
To intercept the launching of an application (\BC{1}),
\sys places a hook in the \cc{execve} system call and maps the customized kernel in that hook (\BC{2}).
In particular,
\sys allocates a new physical memory region and
copies customized kernel code to that region  (\BC{3}). Customized kernel code could either be generated on the fly using a pre-learned kernel profile for the application or could be pre-generated and stored to reduce application launch overhead. We follow the former approach since it only incurs a small one-time launch delay of $~3ms$.
Next,
\sys updates the application's page table entries for kernel text
to point to this new physical memory region  (\BC{4}).
After this point,
the application can only access the customized kernel text
because the original full kernel image will never exist
in its virtual memory space.
Then,
\sys gives control back to the \cc{execve} system call
to handle the loading and linking of the user space, and finally,
we let the execution continue in the user space.

The deployment of the customized kernel and the application is finished at this stage
because kernel switching for each application will be handled automatically
by the regular context switching mechanism as previously discussed.
For instance,~\autoref{fig:switch} shows a case of three applications,
\cc{Apache}, \cc{Php-Fpm} and \cc{MySQL} running in a \sys system.
When a processor (CPU) runs \cc{Apache} in the userspace,
because this application's page table maps kernel code
to the kernel customized for \cc{Apache},
the CPU can interact only with the customized kernel and
cannot access the original full kernel code.
When a context switch happens, say, switching to \cc{php-fpm},
the regular context switching mechanism will change
the value of \cc{CR3} register to the page table of \cc{php-fpm}.
By design,
the page table for \cc{php-fpm} will redirect all accesses to
kernel code to the kernel customized for \cc{php-fpm}.
No matter how a context switch happens in the system,
\sys ensures that
an application can only have a view of kernel code that is customized for it, thus reducing the potential kernel attack surface available to any application.

\PP{Sharing system resources}
We design \sys to allow
sharing of system resources such as
hardware devices,
kernel data structures (\eg task\_struct, \etc),
inter-process communication (IPC) mechanisms
(\ie pipes, UNIX domain sockets, mutexes),
shared memory, \etc -- to maximize a system's flexibility and
compatibility.
More specifically, we design \sys to work with
existing container mechanisms (such as Docker)
and this requires the sharing of system resources
while running customized kernels for each container.
\sys achieves this by
\emph{i}) fully sharing kernel data among customized kernels and
\emph{ii}) having the same memory layout for all customized kernel text.

First,
kernel data and its memory space can be shared as is because \sys does not modify
any of it.
In particular,
any resource sharing that does not involve any kernel text
(such as passing of kernel data structures or device access via DMA)
would not be affected by the deployment of \sys.
However, resource sharing that involves kernel text, for instance,
a data structure that refers to the kernel functions such as function pointers (\eg file operations),
would be affected if any of the customized kernel does have
a different memory layout.
To resolve this issue, \sys requires customized kernels
to have the same memory layout as the original kernel.
That is, a customized kernel will be mapped at exactly the same
virtual address space as that of the original kernel
but the customized kernel text (code) will mask parts of the kernel
not required by that specific application.
Although this approach would create some memory overheads
($~8MB$ for each customized kernel),
this allows \sys to enable sharing of function pointers among
multiple customized kernels. As we see (\autoref{sec:mem-effect}) this memory overhead does not
limit how many applications we can run in parallel in practice.

\subsection{Handling Hardware Interrupts}
\label{s:design-interrupt}

\begin{figure}[!t]
\vspace{-\baselineskip}
	\includegraphics[width=\columnwidth]{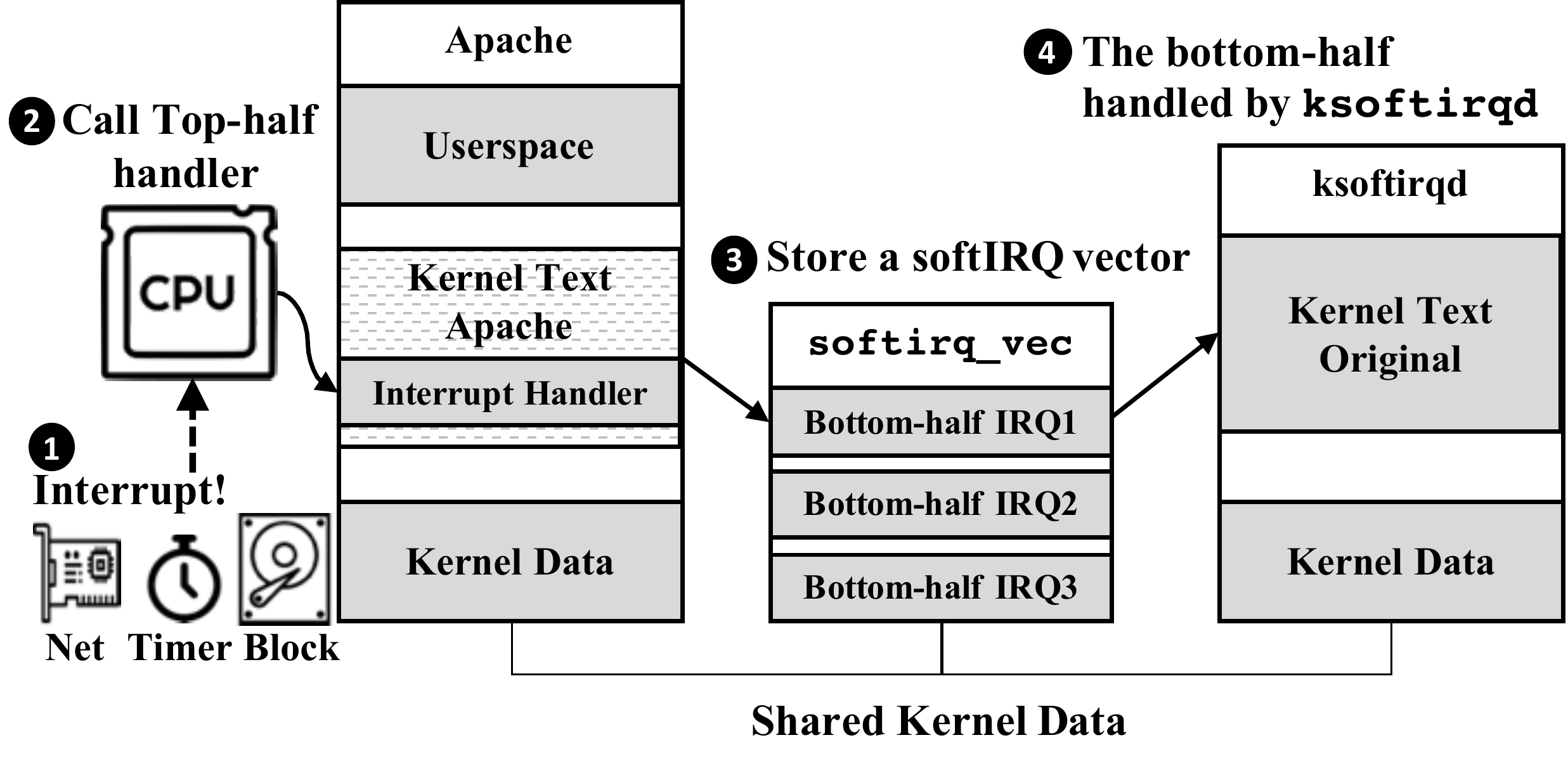}
	\caption{Interrupt handling in \sys.
		1)
		In \sys,
		CPUs that run customized kernel
		can receive a hardware interrupt
		that is not related to an application's execution.
		2)
		To handle such interrupts,
		we include only the part for handling
		the top-half of the interrupt to each customized kernel, and then
		3) this handler will defer this interrupt (to \cc{ksoftirqd}
		by storing a \cc{softirq} vector.
		4) Later,
		\cc{ksoftirqd} will look at the vector and handle
		the bottom-half of the interrupt.
	}
	\label{fig:interrupt}
\vspace{-\baselineskip}
\end{figure}

To handle hardware interrupts
we exploit deferred interrupt handling~\cite{wilcox2003ll}
to keep the customized kernels small.
Hardware interrupts are problematic for customized kernels if
the corresponding handler does not exist in the customized kernel.
Because hardware interrupts caused by the system
could be delivered at any time, even an application that does not utilize
any of the associated hardware
could receive the hardware interrupt requests.
Missing hardware interrupt handlers in customized kernels
will cause such interrupt requests to fail
and the failure to handle such interrupts
could make the system unstable.
One way to work around this issue is
to include all interrupt handling routines in all customized kernels.
However, such an approach would unnecessarily increase the potential attack surface of customized kernels.

% \DL{Also, how do you identify all the top-half handlers?  Is that done through profiling or through a manually constructed whitelist?  Something else?}

\autoref{fig:interrupt} illustrates we handle such interrupts in our framework.
\sys includes only the top-half hardware interrupt handlers
that are compiled into the kernel by whiltelisting (in all customized kernels).
The top-half handlers are smaller because
their job is to transform a hardware interrupt request into a software interrupt request (softirq).
Consequently,
our customized kernels only deal with the top-half of any hardware interrupt and delegate the actual handling of interrupts to
the kernel threads run by \cc{ksoftirqd}.
Hence, when hardware interrupts
(\eg timer, network or block device interrupts) arrive,
our customized kernel will run the top-half handler to
store a softirq vector to delegate
the interrupt handling to \cc{ksoftirqd}.
The bottom-half of the interrupt
will then be handled by \cc{ksoftirqd} when it runs.

\PP{KERNEL0}
To handle the bottom-half of the interrupts,
we run \cc{ksoftirqd} on a general purpose regular kernel
(that includes all parts that the system requires),
that we refer to as \kernelzero.
An example of \kernelzero is a kernel in a distribution's package
without any customization such as \cc{linux-image-4.15.0-39-generic}
in Ubuntu 18.04 LTS.
This kernel not only handles hardware interrupts
but also takes care of system-wide events
such as booting, shutdown, \etc
\kernelzero also serves as
a baseline template for kernel customization
because it contains the entire kernel code required by the system.
In \sys,
customized application kernels are generated by cloning
the kernel \cc{.text} region of \kernelzero
and masking parts of the kernel that are not needed by the application based on the application's kernel profile.

\section{Generating Application Kernel Profiles}
\label{s:kern-speci}
% Design section overview.
% Design requirements
% Each components
% Challenges
% This section is not for just mentioning what we did at here.
% We need to justify our design decision here.
% Present possible alternative approach and describe why our solution is
% working and better.

% outline
% - design
%  - other binary specialization
%  - why they are not enough
%  - our goals: absolute minimum, no neighbor instructions
% - tracing setup
% - segmentation
% - invariants, different code paths therefore re-run
% - block vs symbol
% - compatibility with other approaches

Application kernel profiles identify which parts of the kernel code are
used by an application and which parts are not.
We use 'granularity' as a unit to quantify the precision of the specialization profiles.
The following granularity levels are used in the paper:
\ci \emph{Basic block level}: Basic block is a set of instructions that are always
executed as a unit without any jumps or branches in between.
The CPU either executes all the instructions in the basic block or executes none.
This makes it one of the most precise levels of specialization~\cite{Face-Change-DSN14}.
\cii \emph{Symbol level}: Kernel interfaces are exported as symbols using
\cc{EXPORT_SYMBOL()}. At this level of granularity all the instructions that make up the
interface are included in the profile even if only certain code paths of the
interface are actually being used.
\ciii \emph{System call level}: Given that syscalls are the main interface through which
user space applications interact with the kernel, we can enumerate the syscalls that are
used by the application and eliminate those that aren't~\cite{DBLP:conf/lisa/CowanBKPWG00,DBLP:conf/uss/WrightCSMK02}.
\civ \emph{Page level}: When a binary is loaded into memory, the (memory) smallest unit that can
be referred to is a page (4\KB or more). Tracing methods that track instruction that are being executed
in the memory can only do so on a page level~\cite{kasr}. This will result in an entire page of instructions
being included in the profile even if only a single instruction was run from that page.
\textit{(v)} \emph{Feature level}: The kernel configuration system, \cc{kconfig}, determines
kernel features that need to be built into the kernel based on the state of certain configuration
options. With this code can be included or eliminated only at a feature level, producing a
much coarser and was used in~\cite{kurmus}.
\sys is flexible enough to use application kernel profiles created using 
different profiling granularities 
in order to deploy customized kernels. We present and evaluate one trace-based (\dkut) 
and one syscall-based approach (\skut)
to demonstrate application kernel profile generation.

\subsection{\dkut: Dynamic Instruction Tracing}
\label{s:dkut}
\begin{figure}[ht]
    \includegraphics[width=\columnwidth,keepaspectratio]{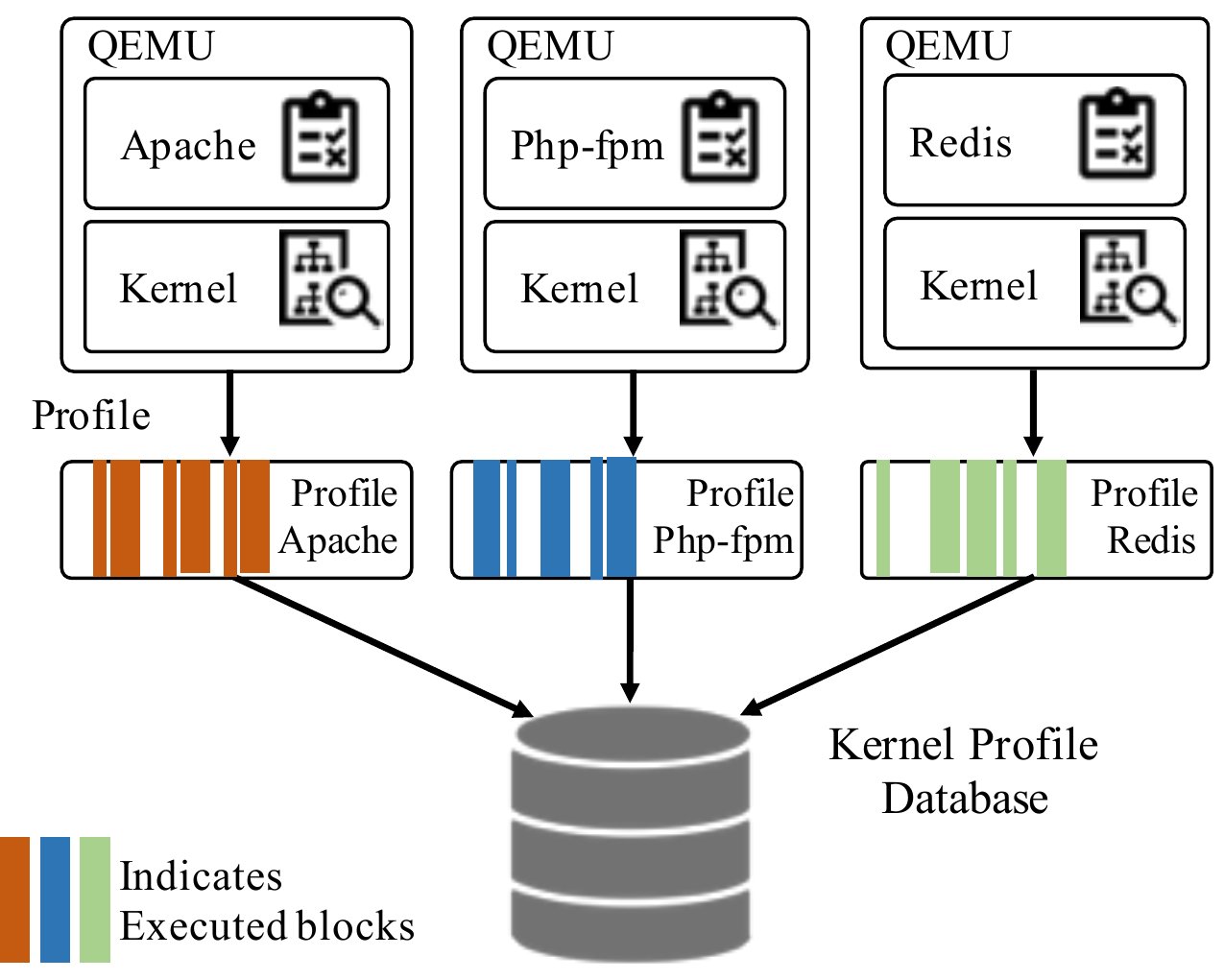}
    \caption{
        \dkut Architecture.
        This figure depicts
        the offline per-application kernel tracing.
        While running each application with test-cases that represents
        application's use cases for kernel,
        \sys trace the kernel and store
        the information for executed basic blocks during the tracing
        into kernel profile database.
    }
    \label{fig:kut}
\end{figure}
\label{sec:dkut}
\dkut (see \autoref{fig:kut}) profiles the kernel at a basic-block level granularity. This is in contrast to KASR~\cite{kasr},
a recently proposed approach for specializing the Linux kernel
binary, that profiles the kernel usage of applications at page level ($4\KB$ default page size) granularity.  When profiled at page level granularity, unused instructions present in the neighborhood of used instructions within the $4\KB$ range also get included in the final in-memory kernel binary.

\PP{Tracing setup.} We use QEMU\cite{qemu}, a full system
emulator to run the user application along with the vanilla kernel
we need to trace. With QEMU we trace every instruction that
the systems executes using the program counter (pc) register by utilizing the \textit{exec_tb_block}.
This basically helps us capture the addresses of instructions that are being \emph{executed}.

\PP{Segmentation.} The trace from above includes instructions from
(1) System boot (2) Applications Runs and (3) System shutdown.
The instructions from (1) and (3) are clearly not required by
the application. To separate this out from the trace we
use the \textit{mmap} syscall to taint/touch a specific memory address
right before the application is run and right after the
application is terminated. This helps us mark the start and end
of our application and helps filter
out the boot/shutdown instructions from the traces.

\PP{Background processes / daemons.} During a typical kernel boot it starts
a bunch of background processes/daemons that provide useful
services. These processes are always running with the application process. They can add noise to the traces that we obtain. To filter these out we use a custom \textit{init} script that only starts daemons required by the application
we are interested in. This eliminates noise from unnecessary background processes  in the traces.

\PP{Invariants in execution.}
Kernel behavior is not deterministic when running an application.
Execution paths in the kernel could change because
not all values are invariants (\eg network conditions, CPU frequencies and time).
Besides, QEMU drops trace events occasionally due to the full buffer.
Therefore to capture all the possible code paths
a kernel might take for a given application,
we repeat the tracing process multiple times.
In our experiments we observed that it takes between 10 to 15 runs before
the code paths stabilize and the trace can be used with confidence as
no new code paths are added for a given set of input.
This need for multiple trace runs has also been reported in prior work~\cite{kurmus,kasr}.

\PP{Granularity.} QEMU produces a trace with \emph{basic block}
level granularity.
% \sys can be configured to used the basic
% block level trace as is or resolve this trace to a binary
% symbol and include it in the final trace.
We can post-process basic-block level traces to obtain symbol-level trace
by including the entire symbol corresponding to each of the blocks.
Symbol-level trace requires less number of runs because
the post-processing makes the trace more inclusive. In our experiments we observe that even at this symbol granularity the trace obtained is much smaller than the ones produced by a page level trace used by KASR~\cite{kasr}, see \autoref{tab:reduction}.

\PP{Compatibility with other specialization techniques.}
\dkut only requires the final bootable kernel binary to produce
a trace. Thus a kernel produced as a result of any other
specialization technique can be further specialized with our tool. Thus \dkut can complement and take advantage of other profiling techniques.

\subsection{\skut: Syscall Analysis}
\label{sec:syscallanalysis}

% However, \sys is independent of the code profiling and
% how to generate the profile is up to users' choice.
% To show that \sys can work seamlessly with different kinds of code profiling techniques,
%
\skut is a syscall-based analysis technique
to increase the reliability of the application kernel profile.
We track all possible functions that a system call can use
for all such calls made by an application.
This list of functions is built by analyzing the register transfer language (RTL)
dumped by GCC when compiling the kernel with \cc{-fdump-rtl-expand}.
We obtain the approximate list of system calls that an application issues by using \cc{strace} 
to intercept and record all system calls made in that execution context.
This approach does not guarantee a complete list of system calls if some
system calls are not triggered during tracing.
% \AU{enhance the reliability e.g. including exception handling code}
We then expand the application profile by combining functions called by possible system
calls with the original profile.
In our experiments, by using this technique, we increase 
the coverage of the profile generated by \dkut at symbol granularity.
The result in \autoref{s:sec-eval} shows that more than $82\%$ of the kernel can be reduced
and $73\%$ of the CVEs are mitigated when including functions used by
system calls in the profile.

\section{Implementation}
\label{s:impl}
% Kernel version
% System setup - CPU, RAM, etc.
% Source line of code
% Tracer
% Kernel source lines
% xxx

\begin{table}
\vspace{-\baselineskip}
	\caption{Source lines of code (SLOC) of the components of \sys and \dkut.}
	\label{tab:loc}
	\centering
	\small
	\begin{tabular}{lr@{~}l}
		\toprule
		\textbf{Component}                & \multicolumn{2}{l}{\textbf{Lines of code}}                 \\
		\midrule
		\dkut - QEMU                       & 3                                          & LoC of C      \\
		\dkut - Scripts                    & 272                                        & LoC of Python \\
		\skut - Make                 & 1
		            & LoC of Makefile \\
		\skut - Scripts                 & 26
		            & LoC of Python \\
		 \skut - Scripts                 & 96
		            & LoC of Golang \\
		\sys - Linux Kernel \linuxversion & 477                                        & LoC of C      \\
		\bottomrule
	\end{tabular}
\vspace{-\baselineskip}
\end{table}

We implement \sys on Linux 4.4.1~\footnote{
We choose an older kernel to
demonstrate \sys's capability in
reducing vulnerabilities by listing
affected CVEs.
Note that \sys can be applied to the newest kernels as well
to remove potential vulnerabilities.
}
running on Intel Core i7-8086K (4.00~GHz) CPU.
We use the \cc{procfs} interface
to provide an application identifier and
the corresponding application kernel profile
to the respective application.
We generate application kernel profiles in an offline step using
\dkut and \skut tools
(see \autoref{s:kern-speci}).
We hook the \cc{execve} syscall so that
when it is invoked to launch an application
(that has a corresponding application kernel profile)
we generate and deploy a specialized kernel as described in \autoref{s:multik}.
%
%In the specialized kernel,
%we mask the unused kernel text region
%with a one-byte opcode \cc{0xcc}, instruction \cc{int3},
%to generate an immediate fault
%when an execution hits any of the removed parts (seel below for details).
%
\autoref{tab:loc} lists the total amount of code for \sys.
We built \sys with $477$ lines of C,
\dkut with $275$ lines of code and
\skut with $123$ lines of code.

Our customized kernels are masked 
in a special manner -- by overwriting them with a special one-byte instruction sequence \cc{0xcc} (instruction \cc{int3}).
This acts a fall-back mechanism for detecting (and reporting)
when an application tries to execute code not available in its customized kernel. 
%
%We mask the unused kernel parts by overwriting them with a special one-byte instruction sequence \cc{0xcc} (instruction \cc{int3}).
%
We choose \cc{int3} not only because this instruction could raise
a software interrupt (that our kernel can intercept to thwart unexpected execution)
but also because it is a one-byte instruction whose semantics cannot be changed by
arbitrary code execution resulting from attacks. 
%(\eg \hl{launching a ROP attack in the middle of an instruction}).
Note that other techniques\cite{Face-Change-DSN14}, overwrite the kernel binary with 
a sequence of \cc{UD2} (two-byte) instructions, \cc{0x0f 0x0b}. 
This opcode can be misinterpreted by reading
it as \cc{0x0b 0x0f}. In such cases, it would not raise an interrupt.
% \DL{Give reader some context.  How does this compare to other systems?  Is it better and in
% what way?  or is it comparable?}
On the other hand, using the \cc{int3} instruction,
we can detect unexpected execution in a more reliable fashion.
%and
%even user program (that requested the execution)
%crash in a benign way.

For instance, consider a simple case when an application's requested kernel execution
hits a masked function or masked basic blocks
(\eg a branch not taken or a system call not used during the tracing).
In such cases, the execution will hit the \cc{int3} instruction immediately.
The kernel knows that the code pointed to (by the instruction pointer for the software
interrupt) is missing. At that point, one can choose to kill the process or follow
other strategies\footnote{Depends on the policies picked by the system designer.}.

\section{Evaluation}
\label{s:eval}

We evaluated \sys to answer the following questions:
\squishlist
\item How much attack surface can \sys reduce (\autoref{s:sec-eval})?
\item How effective is \sys at eliminating CVEs (\autoref{s:case-study})?
\item How long does it take to generate a kernel profile for an application
(\autoref{s:tracing-eval})?
\item What is the effect of \sys on the
system's runtime performance (\autoref{subsec:perfeval})?
\squishend

% evaluation metrics
% - Performance Benchmarks
%   - micro benchmark, application benchmark, benchmark suite
% - Transparency
%   - does it require understanding kernel internals
%   - can it be applied late stage
% - Security
%   - attack surface reduction
%     - in terms of size and granularity / syscalls / ksyms / other APIs
%   - in memory CVE reduction
%   - safety against root kits
\PP{Evaluation setup}
We specialize application kernels at three different granularity:
\ca block and \cb symbol (both with \dkut) and
\ccc syscall (with \skut) (\autoref{s:kern-speci}).
We refer block, symbol and syscall granularity to
\textbf{B}, \textbf{S}, and \textbf{SC}, respectively,
in the rest of the paper.
We performed all experiments in
a KVM virtual machine with 2 vcpus and 8G~RAM running on
Intel(R) Core(TM) i7-8086K CPU @ 4.00GHz.
Note that we use KVM for the convenience of testing,
and \sys does not requires a virtual machine to run
specialized kernels.
For specializing kernels to applications,
we choose Apache, STREAM, perf, Redis, and Gzip.

\subsection{Attack Surface Reduction}
\label{s:sec-eval}

%\begin{figure*}[h]
%	\includegraphics[width=\textwidth, keepaspectratio]
%	{figures/reduction}
%	\caption{Percentage of Attack Surface Reduction compared to the Vanilla Kernel.
%		\granularitycaption
%		}
%	\label{fig:reduction}
%\end{figure*}

\begin{table*}[ht]
\vspace{-\baselineskip}
	\caption{Attack Surface Reduction in comparison with the Vanilla Kernel.
		Text refers to eliminated kernel executable text.
		Full and partial symbol refer to kernel functions that are fully and partially eliminated respectively. \granularitycaption}
	\label{tab:reduction}
	\footnotesize
\begin{tabularx}{\textwidth}{X||c|c|c|c|c|c|c|c|c}
	%\toprule
	\Xhline{2\arrayrulewidth}
	{\bf Specialized Kernel} &
	{\bf Apache-B}           &
	{\bf Apache-S}           &
	{\bf Apache-SC}          &
	{\bf STREAM-B}           &
	{\bf STREAM-S}           &
	{\bf STREAM-SC}          &
	{\bf perf-B}             &
	{\bf perf-S}             &
	{\bf perf-SC}                                                                                                      \\
	\hline

    \bf Text (code)                 & 93.68\% & 88.88\% & 82.25\% & 97.79\% & 95.16\% & 87.64\% & 96.33\% & 93.46\% & 86.57\% \\
	\bf Full Symbol          & 91.87\% & 91.95\% & 85.78\% & 96.65\% & 96.67\% & 89.50\% & 94.78\% & 94.85\% & 88.46\% \\
	\bf Partial Symbol       & 99.78\% & 91.95\% & 85.78\% & 99.93\% & 96.67\% & 89.50\% & 99.84\% & 94.85\% & 88.46\% \\
% 	\bf ROPGadget            & 69.06\% & 65.63\% & 59.66\% & 72.32\% & 70.34\% & 63.55\% & 70.99\% & 68.71\% & 62.57\% \\
% 	\bf ROPPER               & 72.25\% & 68.81\% & 63.08\% & 75.69\% & 73.62\% & 67.10\% & 74.33\% & 72.06\% & 66.15\% \\

	%\bottomrule
	\Xhline{2\arrayrulewidth}
\end{tabularx}

\end{table*}

\begin{table*}[ht]
	\caption{List of CVEs tested in our security evaluation.
	Category numbers are referring to:
	{\bf V}: the entire vulnerability is removed,
	{\bf P}: parts of vulnerability is removed,
	{\bf E}: vulnerability still exists.
	Effects are referring to:
	{\bf DoS}: denial-of-service,
	{\bf Leak}: information leak,
	{\bf Priv}: privilege escalation.
	\granularitycaption
	}
	\label{tab:cves}
	\footnotesize
\newcommand{\cve}[3]{ #1 & #2 & #3}
\begin{tabular}[ht]{lp{7.5cm}c||c|c|c||c|c|c||c|c|c}

	\Xhline{2\arrayrulewidth}
	\multirow{2}{*}{\bf CVE Number}                                                                                                             &
	\multirow{2}{*}{\bf Description}                                                                                                            &
	\multirow{2}{*}{\bf Effect}                                                                                                                 &
	\multicolumn{3}{c||}{\bf Apache}                                                                                                            &
	\multicolumn{3}{c||}{\bf STREAM}                                                                                                            &
	\multicolumn{3}{c}{\bf perf}
	\\ \cline{4-12}
	                                                                                                                                            &   &   & \bf S & \bf B & \bf SC & \bf S & \bf B & \bf SC & \bf S & \bf B & \bf SC \\
	\Xhline{\arrayrulewidth}                                                               %                                                     &   &   &       &       &        &       &       &        &                        \\
	\cve{CVE-2018-11508}{An information leak in \cc{compat_get_timex()}}{\bf Leak}                                                              & V & V & V     & V     & V      & V     & V     & V      & V                      \\
	\cve{CVE-2018-10881}{An out-of-bound access in \cc{ext4_get_group_info()}}{\bf DoS}                                                         & V & V & V     & V     & V      & V     & V     & V      & V                      \\
	\cve{CVE-2018-10880}{A stack-out-of-bounds write in \cc{ext4_update_inline_data()}}{\bf DoS}                                                & V & V & V     & V     & V      & V     & V     & V      & V                      \\
	\cve{CVE-2018-10879}{A use-after-free bug in \cc{ext4_xattr_set_entry()}} {\bf DoS}                                                         & V & V & V     & V     & V      & V     & V     & V      & V                      \\
	\cve{CVE-2018-10675}{A use-after-free bug in \cc{do_get_mempolicy()}} {\bf DoS}                                                             & V & V & V     & V     & V      & V     & V     & V      & V                      \\
	\cve{CVE-2018-7480}{A double-free bug in \cc{blkcg_init_queue()}}{\bf DoS}                                                                  & V & V & V     & V     & V      & V     & V     & V      & V                      \\
	\cve{CVE-2018-6927}{An integer overflow bug in \cc{futex_requeue()}} {\bf DoS}                                                              & E & E & E     & V     & V      & E     & V     & V      & V                      \\
	\cve{CVE-2018-1120}{A flaw in \cc{proc_pid_cmdline_read()}}{\bf DoS}                                                                        & V & V & V     & V     & V      & V     & V     & V      & V                      \\
	\cve{CVE-2017-18270}{A flaw in \cc{key_alloc()}}{\bf DoS}                                                                                   & V & V & E     & V     & V      & E     & V     & V      & E                      \\
	\cve{CVE-2017-18255}{An integer overflow in \cc{perf_cpu_time_max_percent_handler()}}{\bf DoS}                                              & V & V & V     & V     & V      & V     & V     & V      & V                      \\
	\cve{CVE-2017-18208}{A flaw in \cc{madvise_willneed()}}{\bf DoS}                                                                            & V & V & V     & V     & V      & V     & V     & V      & V                      \\
	\cve{CVE-2017-18203}{A race condition between \cc{dm_get_from_kobject()} and \cc{__dm_destory()}}{\bf DoS}                                  & V & V & V     & V     & V      & V     & V     & V      & V                      \\
	\cve{CVE-2017-18174}{A double free in \cc{pinctl_unregister()} called by \cc{amd_gpi_remove()}}{\bf DoS}                                    & V & V & V     & V     & V      & V     & V     & V      & V                      \\
	\cve{CVE-2017-18079}{A null pointer deference in {\scriptsize \cc{i8042_interrupt()}, \cc{i8042_start()}, and \cc{i8042_stop()}}}{\bf DoS}  & V & V & V     & V     & V      & V     & V     & V      & V                      \\
	\cve{CVE-2017-17807}{Lack of permission check in {\scriptsize \cc{request_key_and_link()} and \cc{construct_get_dest_keyring()}}}{\bf Priv} & V & V & E     & V     & V      & E     & V     & V      & E                      \\
	\cve{CVE-2017-17806}{Lack of validation in \cc{hmac_create()} and \cc{shash_no_set_key()}}{\bf DoS}                                         & V & V & V     & V     & V      & V     & V     & V      & V                      \\
	\cve{CVE-2017-17053}{A use-after-free bug in \cc{init_new_context()}}{\bf DoS}                                                              & E & E & E     & V     & V      & E     & E     & E      & E                      \\
	\cve{CVE-2017-17052}{A use-after-free bug in \cc{mm_init()}}{\bf DoS}                                                                       & E & E & E     & V     & V      & E     & E     & E      & E                      \\
	\cve{CVE-2017-15129}{A use-after-free bug in \cc{get_net_ns_by_id()}}{\bf DoS}                                                              & V & V & V     & V     & V      & V     & V     & V      & V                      \\
	\cve{CVE-2017-2618}{Lack of input check in \cc{selinux_setprocattr()}}{\bf DoS}                                                             & V & V & V     & V     & V      & V     & V     & V      & V                      \\
	\cve{CVE-2016-0723}{A use-after-free in \cc{tty_ioctl()}}{\bf DoS}                                                                          & E & E & E     & E     & E      & E     & E     & E      & E                      \\
	\cve{CVE-2015-8709}{A flaw in \cc{ptrace_has_cap()} and \cc{__ptrace_may_access()}}{\bf Priv}                                               & P & P & P     & V     & V      & V     & V     & V      & V                      \\
	\cve{CVE-2015-5327}{A out-of-bound access in \cc{x509_decode_time()}}{\bf DoS}                                                              & V & V & V     & V     & V      & V     & V     & V      & V                      \\

	\Xhline{2\arrayrulewidth}
	                                                                                                                                            &\end{tabular}

\vspace{-\baselineskip}
\end{table*}

We first tackle the question of
how much kernel attack surface \sys can reduce.
We do this by measuring
how much kernel text (code) is reduced by specialization.
\autoref{tab:reduction} shows
the percentage of kernel text reduction w.r.t.
the vanilla kernel,
and {\bf B}, {\bf S}, and {\bf SC}
indicate the granularity of specialization.
%
%
% We also use two open-source tools, ROPgadget~\cite{ROPgadget} and
% ROPPER \footnote{\url{https://github.com/sashs/Ropper}}, to measure the unique ROP gadgets found.
%
%
%The reduction for Apache is slightly less than the other two because Apache uses
%more OS services such as files and networks. <-- why ?
%

The first row shows the percentage of reduced kernel code.
The reduction for each application is
depending on the application's kernel usage.
With the block level granularity,
\sys can reduce $93.68\%$ of the kernel text from
Apache (I/O intensive) and
$97.79\%$ of the kernel text from STREAM

More than $82\%$ of the kernel text can still be removed even when the
granularity is coarse such as symbol and syscall.
Our work outperforms KASR~\cite{kasr} and Tailor~\cite{kurmus}, which can reduce $64\%$ and $54\%$ of kernel code respectively.
The second and third rows present the fully and partially removed kernel functions
respectively.
Because a block is smaller than a function body, we
remove parts of functions.
With block granularity, more then $99\%$ of the text is excluded.

We observed that the implementation of \textit{all} system calls in the Linux kernel 
\footnote{We count all functions called by system call entry functions.}
only takes up $14\%$ of the text. Restricting access to system calls (\eg AppArmor~\cite{DBLP:conf/lisa/CowanBKPWG00}, seccomp-bpf~\cite{DBLP:conf/usenix/KimZ13})
%reduction of the system call implementations
%(as implemented in AppArmor~\cite{DBLP:conf/lisa/CowanBKPWG00}) 
would \ca not remove any code (leaving the kernels vulnerable) and 
\cb have a lesser impact than the 
techniques discussed in this paper. This shows the limitations of only 
focusing on the whitelisting of system calls.

\subsection{CVE Case-Study}
\label{s:case-study}
We analyze all CVEs present in Linux \linuxversion by looking at the patch for each one
and detecting which functions are vulnerable. The kernel is compiled with
configuration 4.4.0-87-generic. A number of vulnerabilities are excluded because
they are not present in the kernel binary since they target loadable kernel
modules and we do not load modules. We find that 23 of 72 CVEs exist in the kernel binary. A CVE
might involve multiple functions.

We separate the results into three
categories to indicate the different levels of mitigation: (1) \textbf{V} refers
to the case where all functions associated with a vulnerability are removed (2) \textbf{P}
refers to case where some functions associated with a vulnerability are removed, and (3)
\textbf{E} refers to case where no functions associated with a vulnerability are
removed.
\autoref{tab:cves} shows the result of each of the CVEs.
On average, $20.3$ (out of $23$) CVEs are mitigated
for both block and symbol granularity on average and $17$ out of $23$ CVEs are mitigated for the system-call granularity.

If a CVE is located in a popular code path, it is more likely that an application
exercises it during the offline profiling phrase.
Therefore, such CVEs (\ie \textit{CVE-2017-17053} and
\textit{CVE-2016-0723}) are likely to remain for all applications.
In other words, if a CVE (\eg \textit{CVE-2018-11508} and \textit{CVE-2017-15129})
is on a unpopular code path, there is a good chance to remove it.
\autoref{fig:cve-example} is an example of a CVE that
we remove the vulnerability by partically removing
one of the function required to form an exploit chain.
The vulnerability is about that
attackers can create a
malicious user namespace and wait for a root process to enter \cc{ptrace_has_cap()}
to gain the root privilege. To exploit the CVE,
the attackers need two functions, \cc{ptrace_has_cap()} and
\cc{__ptrace_may_access()}. Hence, removing one of these
functions can mitigate this CVE.

%%%%%%%%%%%%%%%%%%%%%%%%%%%%%%%%%%%%%%%%%%%%%%%%%%%%%%%%%%%%%
% \begin{figure}[hb]
    % \vspace{-1\baselineskip}
    % \rule{\columnwidth}{0.4pt}
    % \scriptsize
    % \input{code/example.c}
    % %\vspace{-1\baselineskip}
    % \rule{\columnwidth}{0.4pt}
    % %\vspace{-1\baselineskip}
    % \caption{In \textit{CVE-2015-8709},
        % \cc{__ptrace_may_access()} calls \cc{ptrace_has_cap()} to
        % check if the process has the capability to access an user namespace.
        % The code is simplified.
    % }
    % \label{fig:cve-example}
    % \vspace{-\baselineskip}
% \end{figure}
%%%%%%%%%%%%%%%%%%%%%%%%%%%%%%%%%%%%%%%%%%%%%%%%%%%%%%%%%%%%%
\newenvironment{code}{\captionsetup{type=listing}}{}
\begin{code}
\scriptsize
\begin{minted}[mathescape,
               linenos,
               numbersep=5pt,
               % gobble=2,
               frame=lines,
               framesep=2mm]{csharp}
static int
ptrace_has_cap(struct user_namespace *ns, unsigned int mode) {
    //...
    return has_ns_capability(current, ns, CAP_SYS_PTRACE);
}
static int
__ptrace_may_access(struct task_struct *task, unsigned int mode) {
    //...
    if (ptrace_has_cap(tcred->user_ns, mode))
    goto ok;
    //...
}
\end{minted}
\caption{In \textit{CVE-2015-8709},
    \cc{__ptrace_may_access()} calls \cc{ptrace_has_cap()} to
    check if the process has the capability to access an user namespace.
    The code is simplified.
}
\label{fig:cve-example}
% \vspace{-\baselineskip}
\end{code}
%%%%%%%%%%%%%%%%%%%%%%%%%%%%%%%%%%%%%%%%%%%%%%%%%%%%%%%%%%%%%
%
We elaborate each category using a CVE for the Apache web server as an example
with \textbf{B}(Block) granularity.

% \YJ{Show how we determine P, and give an example code.
% We actually look at the patch/fix and then detect which were affected by
% vulnerabilities, and then determine whether vuln exists in our kernel
% or not if that part exists (or not).}
% \YJ{Also discuss about the differece with AppArmor,
% e.g., AppArmor restricts calling of a syscall.
% In the kernel space, all the code are still available,
% that is, attacker still make use of those code
% if syscall contains vuln.}\AU{Done}

\begin{enumerate}
    \item \textbf{V}, the vulnerability is entirely removed.
        \textit{CVE-2017-17807} is a vulnerability resulting from an omitted access-control
        check when adding a key to the current task's default request-keyring via
        the system call, \cc{request_key()}. Two functions, \cc{construct_key_and_link()} and \cc{construct_get_dest_keyring()}, are required to realize the vulnerability.
        Therefore, since both are eliminated, attackers have no way to form the chain in order to exploit this CVE.

    \item \textbf{P}, the vulnerability chain is partially removed.
        \textit{CVE-2015-8709}, depicted in~\autoref{fig:cve-example},
        is a flaw that can allow for a privilege escalation.
        Invoking both functions form the exploit chain: \cc{ptrace_has_cap()} and \cc{__ptrace_may_access()}.
        Because our kenrel specialization partially removes one of the functions (\cc{ptrace_has_cap()}),
        Attackers will no longer be able to exploit this CVE.

    \item \textbf{E}, the vulnerability chain remains.
        \textit{CVE-2017-17052} allows attackers to achieve a `use-after-free'
        because the function \cc{mm_init()} does not null out the member \cc{->exe_file} of a new process's \cc{mm_struct}. Attackers can exploit this as none of the functions have been removed.
\end{enumerate}

% A case of CVE-1 - vulnerability removed
% A case of CVE-2 - vulnerability at the intermediate exploit chain was removed
% A case of CVE-3 - vulnerability resides in kernel-0, but no other kernel
%                   instances has the entire set of vulnerabilities
%                   (requied for successful exploit)
%                   and also have no chance to attack kernel-0.
% A fictional case - similar to case of CVE-3, but attacker can play a trick
%                    with multiple processes; e.g., exploit vulnerability 1
%                    via process A, vulnerability 2 via process B, and
%                    vulnerability 3 via process C,
%                    and obtain privilege escalation. Is it possible?
%                    and can we prevent this?

\subsection{Offline Profile Generation Performance}
\label{s:tracing-eval}
We trace applications for 10 iterations for symbol granularity and 15
for block granularity because we observed that the workload tended to be
stable after this many of iterations.
The profiling time depends on how long the workload runs.
\autoref{tab:tracing} shows the time needed to profile each of the applications.
The time needed for profiling depends the workload. If the workload
(STREAM\cite{stream} in \autoref{tab:tracing}) only takes
a short amount of time to finish, the profiling will be quick and vice versa (Apache and perf in \autoref{tab:tracing}).
\begin{table}[htb]
\vspace{-\baselineskip}
	\caption{Time to complete tracing in seconds.
		\textbf{Symbol} and \textbf{Block} refers to
		\dkut granularity,
		symbol and basic-block respectively.
	}
	\label{tab:tracing}
	\centering
	\small
	\begin{tabular}{l||c|c}
		\toprule
		{\bf Benchmarks} & \textbf{Symbol} & \textbf{Block} \\
		\midrule
		\bf Apache       & 1714.26         & 2492           \\
		\bf STREAM       & 146.77          & 411.98         \\
		\bf perf         & 2139.22         & 2965.15        \\
		\bottomrule
	\end{tabular}
\vspace{-\baselineskip}
\end{table}

\subsection{Performance Evaluation}
\label{subsec:perfeval}
In this section we evaluate \ci application performance by the Apache web server benchmark,
\cii context-switches by perf\footnote{We run the perf command: \cc{perf bench sched all}.} and
\ciii memory bandwidth with STREAM\cite{stream}.
All the experiments were performed in a KVM virtual machine with 2 vcpus and 8G
RAM running on Intel(R) Core(TM) i7-8086K CPU @ 4.00GHz. Positive \% indicates improvements,
and negative \% indicates degradation of the application performance.
% \begin{itemize}
% 	\item With macro benchmarks we evaluate application performance, startup overhead, memory overhead, and interrupt handling overhead
% 	\item With micro benchmarks we evaluate interrupt handling and Softirq forwarding
% \end{itemize}
%
Positive \% indicates improvements,
and negative \% indicates degradation.
We evaluate the performance with
application benchmarks such as
Apache web server,
STREAM~\cite{stream}, a memory microbenchmark,
and \emph{perf}, a scheduling microbenchmark.
Due to the page limits,
we place some of the benchmark results
in Appendix \autoref{sec:appendix-perfeval} including
Redis and Gzip with \dkut at
symbol and block granularity, since the results are similar.

\PP{Apache Web Server}
We ran the Apache web server, version 2.4.25, on a specialized kernel and the
client program on \kernelzero. The client program Apache benchmark sends 100,000 requests
in total with 100 clients concurrently. \autoref{tab:apache-perfeval} shows
that the Apache web server running on specialized kernels (regardless of the
trace granularity) has a very similar performance, in terms of number of requests served per second, compared to when running on a vanilla kernel. The performance is within $0.6\%$ of the baseline.

\begin{table}[b]
\vspace{-\baselineskip}
	\caption{
		Processed HTTP requests per second when running Apache web server
		with and without (vanilla) \sys. \granularitycaption
    Green cells $\rightarrow$ better performance (more requests per second).
    Red cells $\rightarrow$ worse  performance (less requests per second).
	}
	\label{tab:apache-perfeval}
	\centering
	%\begin{tabular}{lccc}
	\begin{tabular}{l||c|c}
		\Xhline{2\arrayrulewidth}
		                   & \metricLast{Req/Sec}                      \\
		\toprule
		\bf Apache-Vanilla & 23401.07             & 0.00 \%            \\
		\bf Apache-B       & 23337.53             & \redcell -0.27\%   \\
		\bf Apache-S       & 23445.03             & \greencell +0.19\% \\
		\bf Apache-SC      & 23536.31             & \greencell +0.58\% \\
		\Xhline{2\arrayrulewidth}
	\end{tabular}
\vspace{-\baselineskip}
\end{table}

%%%%%%%%%%%%%%%%%%%%%%%%%%%%%%%%%%%%%%%%%%%%%%%%%%%%%%%%%%%%%%%%%%%%%%%%%%%%%%%%%%%%%%%%%%%%%%%%%%%%%%%%%%%%

\PP{STREAM}
We evaluate the memory performance with STREAM, version
5.10. STREAM has four metrics including \cc{copy},
\cc{scale}, \cc{add} and \cc{triad}. They refer to the
corresponding kernel vector operations.
\cc{Copy} refers to \cc{a[i] = b[i]}.
\cc{Scale} refers to \cc{a[i] = const*b[i]}.
\cc{Add} refers to \cc{a[i] = b[i]+c[i]}.
\cc{Triad} refers to \cc{a[i] = b[i]+const*c[i]}. \cc{Copy} and \cc{scale} take two memory accesses while \cc{add} and \cc{triad} take three.
\autoref{tab:stream-perfeval} shows that STREAM running on specialized kernels regardless of the granularity have close performance compared to the baseline and the difference is less than $0.5\%$ for all operations.

\begin{table*}
\vspace{-\baselineskip}
	\caption{Average time to complete different kernel vector operations in seconds.
	Copy refers to \cc{a[i] = b[i]}.
	Scale refers to \cc{a[i] = const*b[i]}.
	Add refers to \cc{a[i] = b[i]+c[i]}.
	Triad refers to \cc{a[i] = b[i]+const*c[i]}.
	\granularitycaption
    Green cells $\rightarrow$ better performance (shorter time to finish an operation).
    Red cells $\rightarrow$ worse  performance (shorter time to finish an operation).
	}
	\label{tab:stream-perfeval}
	\centering
	\begin{tabular}{l||c|c|c|c|c|c|c|c}
		\Xhline{2\arrayrulewidth}
		\bf Operation      & \metric{Copy} & \metric{Scale}     & \metric{Add} & \metricLast{Triad}                                                                 \\
		\toprule
		\bf STREAM-Vanilla & 0.008192      & 0.00\%             & 0.008632     & 0.00\%             & 0.011207 & 0.00\%             & 0.011428 & 0.00\%             \\
		\bf STREAM-B       & 0.008201      & \redcell -0.11\%   & 0.008623     & \greencell +0.10\% & 0.011237 & \redcell -0.27\%   & 0.011445 & \redcell -0.15\%   \\
		\bf STREAM-S       & 0.008199      & \redcell -0.09\%   & 0.008628     & \greencell +0.05\% & 0.011201 & \greencell +0.05\% & 0.011418 & \greencell +0.09\% \\
		\bf STREAM-SC      & 0.008178      & \greencell +0.17\% & 0.008597     & \greencell +0.41\% & 0.011179 & \greencell +0.25\% & 0.011406 & \greencell +0.19\% \\
		\bottomrule
	\end{tabular}
\vspace{-\baselineskip}
\end{table*}

%%%%%%%%%%%%%%%%%%%%%%%%%%%%%%%%%%%%%%%%%%%%%%%%%%%%%%%%%%%%%%%%%%%%%%%%%%%%%%%%%%%%%%%%%%%%%%%%%%%%%%%%%%%%

\PP{Perf}
We evaluate context switch overheads with the perf scheduling benchmark that is composed of a
messaging microbenchmark and a piping microbenchmark. The messaging microbenchmark
has 200 senders that dispatch messages through sockets  to 200 receivers concurrently.
The Piping microbenchmark has 1 sender and 1 receiver processes executing 1,000,000
pipe operations. \autoref{tab:perf-perfeval} shows that perf running on specialized kernels
(regardless of the trace granularity) takes the same amount of time to complete the message and pipe tasks when compared to running on vanilla (unmodified) kernels.
The performance difference for both message and pipe tasks is less than $0.6\%$.
\begin{table}[b]
\vspace{-\baselineskip}
	\caption{Time to complete each microbenchmark in seconds.
		Message refers to concurrent sender/receiver microbenchmark.
		Pipe $\rightarrow$ sequential piping between two process.
		\granularitycaption
    Green cells $\rightarrow$ better performance (shorter time to finish benchmark).
    Red cells $\rightarrow$ worse  performance (shorter time to finish benchmark).
	}
	\label{tab:perf-perfeval}
	\centering
	\begin{tabular}{l||c|c|c|c}
		\toprule
		\bf Benchmark    & \metric{Message} & \metricLast{Pipe}                               \\
		\midrule

		\bf perf-Vanilla & 0.190            & 0.00\%             & 10.37 & 0.00\%             \\
		\bf perf-B       & 0.191            & \redcell -0.53\%   & 10.36 & \greencell +0.10\% \\
		\bf perf-S       & 0.189            & \greencell +0.53\% & 10.35 & \greencell +0.19\% \\
		\bf perf-SC      & 0.190            & 0.00\%             & 10.38 & \redcell -0.10\%   \\
		\bottomrule
	\end{tabular}
	\vspace{-\baselineskip}
\end{table}

% \subsubsection{Startup Overhead}
% Kernel0 needs to allocate and mask a piece of memory for the new
% specialized kernel when launching an application. Memory allocation and
% overwriting introduces 3010 $\mu$s delay. It is an one-time effort for
% the application and it is possible for \kernelzero to cache the memory to
% save the startup overhead if \kernelzero knows the information of kernel
% specialization beforehand.

\subsubsection{Memory Effect}
\label{sec:mem-effect}
\begin{figure}[h]
\vspace{-1\baselineskip}
	\includegraphics[width=\columnwidth, keepaspectratio]
	{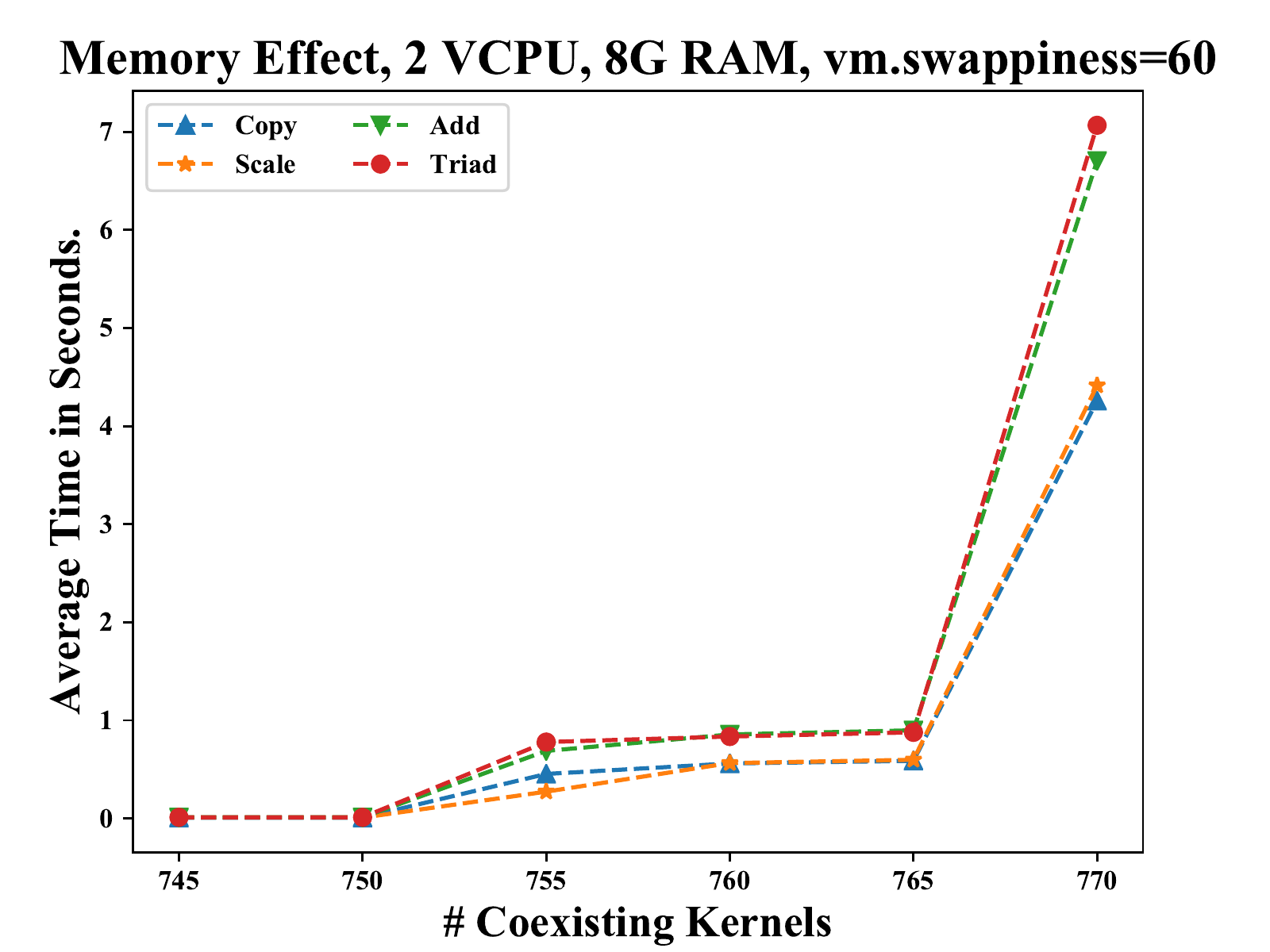}
	\caption{STREAM benchmark with multiple coexisting kernels.}
	\label{fig:memeffect}
%\vspace{-1\baselineskip}
\end{figure}

Every specialized kernel takes up approximately $8MB$ of additional memory space .
Kernel memory is not swappable. Therefore, the system-wide memory
pressure will increase if we keep creating new specialized kernels.
We evaluate the memory pressure by measuring the memory bandwidth with
STREAM~\cite{stream} benchmark Under a higher memory pressure, the system
will swap memory pages in and out more frequently and results in a lower memory bandwidth.
We run STREAM together with multiple specialized kernels on a KVM virtual
machine with 2 vcpus and 8GB running on Intel(R)Core(TM) i7-8086K CPU @ 4.00GHz.
\textit{Swappiness} is a kernel parameter ranging from 0 to 100,
which controls the degree of swapping.
The higher the value is, the more frequently virtual memory swaps.
We conduct the experiment with default value, 60.
In order to exclude the factor of the CPU (\ie busy CPUs cause slower memory
bandwidth),
we run the command \cc{sleep} on
specialized kernels.
\autoref{fig:memeffect} shows that memory operations start to slow down
due to more frequent memory swaps when there are more than \emph{750
coexisting kernels}. To be more specific, operations \cc{Add} and \cc{Triad}
take twice the memory accesses than the others so the time that they need to finish is twice
longer. We also evaluated the memory effect on a machine with 4G RAM running
on Intel(R) Xeon(R) CPU E3-1270 v6 @ 3.80GHz. In this case the impact is observed after $345$ coexisting specialized kernels (refer to
\autoref{sec:app-memeffect}). The results indicate that memory overhead is not an issue for most practical deployments.

\section{Discussion}
\label{s:discus}
\PP{Kernel Specialization for Containers}
The method of \sys for dispatching multiple kernels can
seamlessly assign and run a specialized kernel for
each security domain, \ie container, such as Docker.
This is done by sharing a specialized kernel for multiple applications.
%
%In particular,
%a container image itself can been seen as a pre-defined workload.
%
%These pre-defined workloads enable developers to generate
%the kernel profile with ease.
%
In particular,
we can configure \sys to have
one specialized kernel for set of applications (running in a container)
by profiling kernel traces while running
all such application together.
In our experiment, we integrated \sys with Docker containers trivially,
and show that \sys affects Docker containers' performance by less than
$1\%$ in \autoref{sec:docker}.
This approach can also reduce the memory usage by using fewer kernels.
It aligns well with cloud deployment patterns where
\textit{containers} from different organizations may
share the same hardware. %\hl{\cite{xx}}

\PP{Kernel Page-Table Isolation}
Kernel page-table isolation (KPTI)~\cite{lwnkpti} is a feature that mitigates the
Meltdown~\cite{DBLP:conf/uss/Lipp0G0HFHMKGYH18} security vulnerability.
KPTI uses two page tables for user and kernel
modes respectively so that user-mode applications can only access a limited
range of kernel memory space such as entry/exit functions and interrupt descriptors.
Although KPTI hides the kernel from the user space, it does not mitigate the
vulnerabilities because attackers can still call system calls
with carefully crafted parameters to enter the kernel mode and exploit the system.

\PP{Kernel CFI}
\sys is orthogonal to
other kernel attack surface reduction techniques
such as control-flow integrity (CFI)~\cite{DBLP:conf/ccs/AbadiBEL05}
and can work with them concurrently.
Kernel CFI can indirectly achieve attack surface reductions
by restricting available call/jump targets from
a large number of
control-flow transfer points (that would otherwise serve as
attack surfaces).
Such an approach is orthogonal to \sys
and, in fact, they can both complement each other.
In particular,
\sys can run a kernel reinforced with KCFI,
and \sys can further trim such a kernel
to achieve an overall better attack surface reduction.
The reason that \sys can be seamlessly combined with
such techniques is due to the fact that it incurs
extremely low performance overheads.

\PP{AppArmor and SELinux}
AppArmor~\cite{DBLP:conf/lisa/CowanBKPWG00} and
SELinux~\footnote{\url{https://www.nsa.gov/What-We-Do/Research/SELinux}} are
Linux security modules~\cite{DBLP:conf/uss/WrightCSMK02}
which try to achieve Mandatory Access Control.
In particular,
after identifying the necessary resources and capabilities,
both approaches apply a profile to enable/disable them via white/blacklisting.
The drawbacks of AppArmor and SELinux (compared to \sys) is that
they only remove access to syscalls
that are an entry point to a certain code path
by limiting the POSIX capabilities.
The code is still available to the attacker if she can
bypass~\footnote{\url{https://www.cvedetails.com/cve/CVE-2017-6507}}
this protection.
In \sys we explicitly remove code paths
that are not required by the application, thus preventing the attacker from
accessing it altogether even if other security measures are compromised.
In addition to that, \sys can further reduce code within a system call
if we apply smaller specialization granularity than a symbol,
\eg basic-block granularity.

\PP{Arbitrary Kernel Execution}
If an attacker is able to execute arbitrary code within the kernel
space, \eg by inserting kernel modules, then the attacker can modify the page tables for applications and bypass the kernel view imposed by \sys.
We prohibit kernel module insertion for
specialized kernels. If a kernel module is needed, it can be inserted
from \kernelzero and it is visible to all specialized kernels.

\PP{Kernel Data-Only Attacks}
As the underlying kernel data structures are shared among
all the multiplexed customized kernels,
\sys alone cannot prevent kernel data corruption attacks
(\eg~\cite{dkom,Xiao-SECCOM15, hu:dop, hu:auto-de})
although it can make it harder for attackers to exploit.
However it can be integrated with existing kernel data protection
mechanisms (\eg~\cite{Xiao-SECCOM15}) to improve the overall security of the system.

\section{Related Work}
\label{s:related}

In this section we discuss and compare different kernel reduction approaches. They can be broadly classified into \ci configuration based
specialization, \cii compiler based specialization, \ciii binary specialization, or \civ kernel re-architecture.

\PP{Configuration based specialization.}
The Linux kernel provides the \kconfig mechanism for configuring the kernel.
However, the complexity of \kconfig makes it hard
to tailor a kernel configuration for a given application.
Kurmus \etal~\cite{kurmus} tried to automate \kconfig based kernel customization
by obtaining a runtime kernel trace for a target application,
then mapping the trace back to the source lines
and using source lines along with configuration dependencies for
arriving at an optimal configuration.
While they were able to achieve 50-80\% reduction in kernel size,
their approach still requires some manual effort for creating
predefined blacklists and whitelists of configurations~\cite{lwnkernelconfig}.
Further, when multiple applications need to be run,
their approach creates a customized kernel for
all of the applications together thereby limiting
the effectiveness of the attack surface reduction achieved~\cite{Face-Change-DSN14}.
More recently, LightVM~\cite{lightvm} tries to address
bloat in the kernel by implementing a tool called TinyX
that starts from Linux's \cc{tinyconfig}
and iteratively adds options from a bigger set of
configuration options. This involves maintaining a
manually produced white-list or black-list.
In \sys, everything is automated and requires no manual
intervention.
%\AU{I don't think it reduces isolation, but functionality. the achieve lightvm by either unikernel or tinyconfig-linux. I think it's unfair to compare to dokcer on distro Linux.}

The Linux kernel tinification project~\cite{tinykernel} focuses on reducing
the size of the kernel by making every core kernel feature configurable, thereby
allowing developers to build just the minimum set of features
required to run on embedded devices.
Although configuration-based specialization can be effective,
it remains a manual and tedious process.
\sys completely circumvents the configuration process
and specializes the kernel binary directly.
We are also able to produce a more fine grained reduction
compared to the coarse grained reduction produced by
configuration specialization~\cite{kurmus}.

\PP{Compiler based specialization.}
Modern compilers are much better at code optimization than humans are.
A series of LWN.net articles~\cite{lwngarbagecollection, lwnlto, lwnhammer, lwnaxe}
discusses various cutting edge efforts in compiler and link-time techniques
that are being developed in the Linux community
that can eliminate significant amount of dead code and
perform various other code optimizations.
Most of the work in this area is experimental and does not produce
a working kernel yet.
They exist as out-of-tree patches~\cite{kleenlto14}.
The main challenge in applying these techniques to the Linux kernel
arises out of the complexities in the kernel itself.
Handwritten assembly, non-contiguous layout of functions, \etc do not
make the kernel a good candidate for compiler-based
optimization/specialization as is. It requires manually going through
pieces of code and making careful changes without causing
unexpected side effects.
The LLVM community in recent years has produced a suite of
advanced compiler tools~\cite{allvm}.
The Linux community hasn't been able to take advantage
of these due to its tight coupling with the gcc toolchain,
making it hard to use other compilers like clang~\cite{lwnkernelclang} from the LLVM tool chain to build the kernel.
These are being fixed one patch at a time~\cite{clang-kernel}.
Our approach works with binary instructions and does not depend on
a specific compiler tool chain. Moreover compilers will only be able to 
produce reduction from static analysis methods.
There is no room for reduction based on run time behavior.
We are able to take into consideration the run-time behavior
and make further reductions.

\PP{Binary specialization.}
Binary specialization techniques do not require
reconfiguring or rebuilding the kernel.
They work on the final kernel binary as is.
KASR~\cite{kasr} specializes the kernel binary
using a VM-based approach wherein
they trace all the pages in the memory that are
used by an application -- for a few iterations until the trace doesn't change.
This data is used to mark the unused pages
in the extended page tables as non-executable,
thus making the memory region unavailable to the application.
Face-Change~\cite{gudsn14}, shadow kernels~\cite{chickapsys15} create specialized
kernel text areas for each of the target applications and switch between them
using a hypervisor to support multiple applications running together.
Their performance is limited by the performance of the hypervisor.
Face-Change reported performance overheads approx. $40\%$ for I/O benchmarks and doesn't support multithreading.
\kasr customizes the kernel at a page level granularity. \sys has near native performance while customizing the kernel at basic-block level granularity while supporting multithreading. %Moreover \sys can support different specialization techniques in-addition to binary specialization and can be extended to use different specialization technique for each application.

\PP{Kernel redesign.}
An orthogonal direction to specializing general purposes kernels
for attack surface reduction is to use unikernels or microkernels
that define a completely new architecture.
Unikernels~\cite{unikernels} get rid of protection rings and
have the application code and the kernel in a single ring
to reduce performance overhead arising from  context switches.
But this leaves kernel code that is required for the entire system
including boot and termination available to the application.
Microkernels such as Mach~\cite{dravesmkern92} design the kernel in
a very modular manner such that the kernel TCB is minimal.
This comes at the cost of performance overhead from context switching.
Moreover both these approaches require the application to
be re-built for the respective architectures.
NOOKS~\cite{nookssigops02} redesigns the kernel to
isolate device drives from the kernel core -- to protect
from vulnerable device drivers. It still leaves active a major
part of the kernel (for boot, shutdown and
other OS tasks).
%that are not required by the application view of the kernel.

%\hl{In our customization work we use a binary specialization method, which is able to produce a very precise cut.}

\section{Conclusion}
\label{s:conclusion}
\sys does not aim to provide the same level of isolation as virtual machines(VM).
Instead, it is a framework that runs specialized kernel code without losing
the flexibility of integrating existing security mechanism to defend against different
cyber-threats \eg data-only attacks.
Our evaluation shows that \sys can effectively reduce the kernel 
attack surface while multiplexing multiple commodity/specialized kernels with 
less than $1\%$ overhead. \sys can be easily integrated with container-based 
software deployment to achieve per-container kernels with no changes to the application.

{\normalsize
	\bibliographystyle{unsrt}
	\bibliography{bib,conf}
}

\clearpage
% \theendnotes
\appendix
\section{Appendix}
\label{sec:appendix}

\subsection{Docker Performance}
\label{sec:docker}
\autoref{tab:docker-perfeval} shows the performance evaluation when \sys is integrated with Docker containers.

\subsection{Redis and Gzip Benchmarks}
\label{sec:appendix-perfeval}
\PP{Redis}
We ran the \cc{redis-server} (version 3.2.6) on the specialized kernel and exercised
the \cc{redis-benchmark} with default configuration on \kernelzero. The benchmark sends 10,0000 requests of each Redis command with 50 concurrent clients and measures the number of requests serviced per second.
\autoref{tab:redis-perfeval} shows that the requests per second for each command, and that the Redis server running on both symbol- and block-granularity kernels can achieve performance similar to that running on vanilla kernel for all tested commands ($<+/-1\%$).

\PP{gzip}
We measure time spent in both kernel and user modes to compress a randomly
generated 512MB file with Gzip 1.6.
\autoref{tab:gzip-perfeval} shows that gzip running on
specialized kernels, regardless of the trace granularity,
has similar performance compared to running on vanilla kernel.

\subsection{Memory Effect}
\label{sec:app-memeffect}
\begin{figure}[h]
    \includegraphics[width=\columnwidth, keepaspectratio]
    {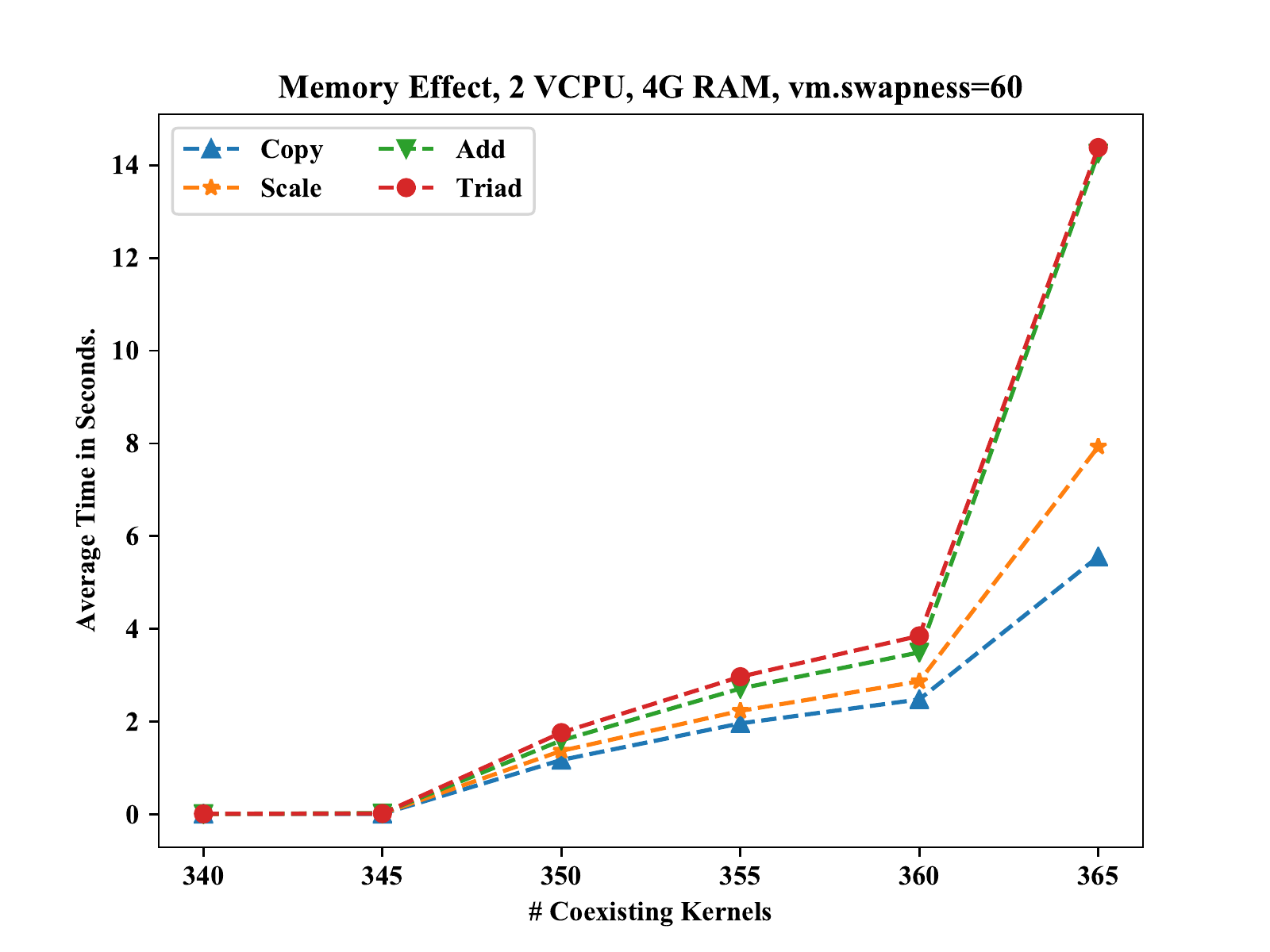}
    \caption{STREAM benchmark with different numbers of coexisting kernels.}
    \label{fig:memeffect-old}
\end{figure}

\autoref{fig:memeffect-old} shows the memory effect with STREAM\cite{stream} in a KVM instance with 2 VCPU and 4G RAM running on Intel(R) Xeon(R) CPU E3-1270 v6 @ 3.80GHz.

\subsection{ROP Gadget Attacks}
We use two open-source tools, ROPgadget~\cite{ROPgadget} and
ROPPER \footnote{\url{https://github.com/sashs/Ropper}}, to measure the unique ROP gadgets found.
\autoref{tbl:rop-reduction} indicates the reduced number of ROP gadget reported by
these two tools. ROPPER can find more gadgets than ROPGadget by $3\%$ to $6\%$.
A reduction in the number of gadgets by itself is not a good 
indicator for ROP attack evaluation because attackers can still 
exploit the system if they can form chains they need from the 
remaining gadgets~\cite{DBLP:conf/ccs/AbadiBEL05}. We report these
reductions (\autoref{tbl:rop-reduction}) as a guide to system designers, but they
should be taken with a grain of salt.
%However, we still report this number as attackers might
%need to spend more effort to exploit the system or be unable to form a critical ROP chain.
\sys can also be coupled with CFI~\cite{DBLP:conf/ccs/AbadiBEL05} 
to improve the protection of the system from ROP attacks.

%%%%%%%%%%%%%%%%%%%%%%%%%%%%%%%%%%%%%%%%%%%%%%%%%%%%%%%%%%%%%%%%%%%%%%%%%%%%%%%%%%%%%%%%%%%%%%%%%%%%%%%%%%%%

%%%%%%%%%%%%%%%%%%%%%%%%%%%%%%%%%%%%%%%%%%%%%%%%%%%%%%%%%%%%%%%%%%%%%%%%%%%%%%%%%%%%%%%%%%%%%%%%%%%%%%%%%%%%

\begin{table}[h!]
    \caption{Time to compress a randomly generated 512MB file in seconds with Gzip. The unit is seconds.
        \textbf{Real} refers to the wall clock time.
        \textbf{User} refers to the amount of CPU time spent in user-mode code
        within the process.
        \textbf{Sys} refers to the amount of CPU time spent in kernel-mode code
        within the process.
        \granularitycaptionWoSc
        Green cells $\rightarrow$ better performance (shorter time to finish compression).
        Red cells $\rightarrow$ worse  performance (longer time to finish compression).
    }
    \label{tab:gzip-perfeval}
    \centering
    \small
    \begin{tabular}{l||c|c|c|c|c}
        \toprule
        {\bf Type} & {\bf Vanilla} & \multicolumn{2}{c|}{\bf gzip-S} & \multicolumn{2}{c}{\bf gzip-B}                             \\
        \midrule
        real       & 10.049        & 10.05                           & \redcell -0.02\%               & 10.088 & \redcell -0.39\% \\
        user       & 9.56          & 9.58                            & \redcell -0.25\%               & 9.648  & \redcell -0.92\% \\
        sys        & 0.512         & 0.504                           & \greencell +1.56\%             & 0.516  & \redcell -0.78\% \\

        \bottomrule
    \end{tabular}
\end{table}

% \begin{table}[h!]
% \caption{
% Time to complete tracing in seconds.
% \granularitycaptionWoSc
% }
% \label{tab:tracing-appendix}
% \centering
% \begin{tabular}{l||c|c}
% \toprule
% {\bf Benchmarks} & {\bf Symbol} & {\bf Block} \\
% \midrule
% gzip             & 664.04       & 999.68      \\
% Redis            & 12710.43     & 17495.36    \\
% \bottomrule
% \end{tabular}
% \end{table}

\begin{table*}[h]
    \caption{Performance comparison between applications in docker containers on vanilla kernel and \sys + \dkut.
        \granularitycaptionWoSc
        For Apache, the higher number, the better performance. For the others, the lower number, the better performance.
            Green cells $\rightarrow$ better performance.
            Red cells $\rightarrow$ worse  performance.
    }
    \label{tab:docker-perfeval}
    \centering
    \begin{tabular}{l||c|c|c|c|c}
        \toprule
        {\bf Application}                & {\bf Vanilla} & \multicolumn{2}{c|}{\bf Symbol} & \multicolumn{2}{c}{\bf Block}                                   \\
        \midrule
        Apache (req per sec)             & 13999.090     & 13988.560                       & \redcell -0.08\%              & 13994.800 & \redcell -0.03\%    \\
        gzip a 512\MB file (sec)         & 11.062        & 11.052                          & \greencell +0.01\%            & 11.022    & \greencell  +0.36\% \\
        perf bench sched messaging (sec) & 0.192         & 0.192                           & 0.00\%                        & 0.193     & \redcell -0.52\%    \\
        perf bench sched pipe (sec)      & 16.063        & 16.098                          & \redcell -0.22\%              & 15.953    & \greencell  +0.68\% \\
        \bottomrule
    \end{tabular}
\end{table*}

\begin{table*}[h]
    \caption{
        Requests per second for each of the Redis commands.
        \granularitycaptionWoSc
        Green cells $\rightarrow$ better performance (more requests per second).
        Red cells $\rightarrow$ worse  performance (less requests per second).
    }
    \label{tab:redis-perfeval}
    \centering
    \begin{tabular}{l||c|c|c|c|c}
        \toprule
        {\bf Benchmarks}  & {\bf Vanilla} & \multicolumn{2}{c|}{\bf Redis-S} & \multicolumn{2}{c}{\bf redis-B}                                  \\
        \midrule
        \cc{PING\_INLINE} & 398406.41     & 396825.38                        & \redcell -0.40\%                & 398406.41 & 0.00\%             \\
        \cc{PING\_BULK}   & 411522.62     & 409836.06                        & \redcell -0.41\%                & 413223.16 & \greencell +0.41\% \\
        \cc{SET}          & 414937.75     & 418410.06                        & \greencell +0.84\%              & 416666.69 & \greencell +0.42\% \\
        \cc{GET}          & 396825.38     & 398406.41                        & \greencell +0.40\%              & 400000    & \greencell +0.80\% \\
        \cc{INCR}         & 416666.69     & 418410.06                        & \greencell +0.42\%              & 414937.75 & \redcell -0.41\%   \\
        \cc{LPUSH}        & 427350.44     & 429184.56                        & \greencell +0.43\%              & 425531.91 & \redcell  -0.43\%  \\
        \cc{RPUSH}        & 416666.69     & 416666.69                        & 0.00\%                          & 414937.75 & \redcell -0.41\%   \\
        \cc{LPOP}         & 418410.06     & 420168.06                        & \greencell +0.42\%              & 420168.06 & \greencell +0.42\% \\
        \cc{RPOP}         & 413223.16     & 416666.69                        & \greencell +0.83\%              & 413223.16 & 0.00\%             \\
        \cc{SADD}         & 414937.75     & 418410.06                        & \greencell +0.84\%              & 418410.06 & \greencell +0.84\% \\
        \cc{SPOP}         & 409836.06     & 409836.06                        & 0.00\%                          & 408163.25 & \redcell -0.41\%   \\
        \cc{LPUSH}        & 421940.94     & 421940.94                        & 0.00\%                          & 423728.81 & \greencell +0.42\% \\
        \cc{LRANGE\_100}  & 170068.03     & 170940.17                        & \greencell +0.51\%              & 169491.53 & \redcell -0.34\%   \\
        \cc{LRANGE\_300}  & 41493.77      & 41101.52                         & \redcell  -0.95\%               & 41425.02  & \redcell -0.17\%   \\
        \cc{LRANGE\_500}  & 28490.03      & 28409.09                         & \redcell  -0.28\%               & 28522.53  & \greencell +0.11\% \\
        \cc{LRANGE\_600}  & 21649.71      & 21523.89                         & \redcell  -0.58\%               & 21454.62  & \redcell  -0.90\%  \\
        \cc{MSET}         & 311526.47     & 310559                           & \redcell  -0.31\%               & 311526.47 & 0.00\%             \\
        \midrule
        \textbf{Geo Mean} & 241971.16     & 242334.95                        & \greencell +0.10\%              & 242135.83 & \greencell +0.02\% \\
        \bottomrule
    \end{tabular}
\end{table*}

\begin{table*}[ht]
    \caption{ROP gadget reduction in comparison with the Vanilla Kernel. We measure
    the number of unique ROP gadgets and show the percentage compared to the
    vanilla kernel. \granularitycaption}
    \label{tbl:rop-reduction}
    \footnotesize
    \begin{tabularx}{\textwidth}{X||c|c|c|c|c|c|c|c|c}
        %\toprule
        \Xhline{2\arrayrulewidth}
        {\bf Specialized Kernel} &
        {\bf Apache-B}           &
        {\bf Apache-S}           &
        {\bf Apache-SC}          &
        {\bf STREAM-B}           &
        {\bf STREAM-S}           &
        {\bf STREAM-SC}          &
        {\bf perf-B}             &
        {\bf perf-S}             &
        {\bf perf-SC}                                                                                                      \\
        \hline

            \bf ROPGadget            & 69.06\% & 65.63\% & 59.66\% & 72.32\% & 70.34\% & 63.55\% & 70.99\% & 68.71\% & 62.57\% \\
            \bf ROPPER               & 72.25\% & 68.81\% & 63.08\% & 75.69\% & 73.62\% & 67.10\% & 74.33\% & 72.06\% & 66.15\% \\

        %\bottomrule
        \Xhline{2\arrayrulewidth}
    \end{tabularx}

\end{table*}

\end{document}